\documentclass[letterpaper,twocolumn,10pt]{article}
\usepackage{zhanggroup}

\usepackage{tikz}
\usepackage{amsfonts}
\usepackage{xspace} 
\usepackage{amsmath}
\usepackage{mathrsfs}
\usepackage{amsthm}
\usepackage{subcaption}
\usepackage{booktabs}
\usepackage{multirow}
\usepackage{hyperref}
\usepackage{pifont}
\usepackage{makecell}
\usepackage{stfloats}
\newcommand{\mypara}[1]{\noindent{\bf {#1}.}}

\newcommand{\Graph}{G}
\newcommand{\Subgraph}{g}
\newcommand{\NodeSet}{V}
\newcommand{\EdgeSet}{E}

\newcommand{\TargetNode}{v}

\newcommand{\Radius}{k}
\newcommand{\Neighbors}{N^{k}(v)}

\newcommand{\post}{posterior-only\xspace}
\newcommand{\comb}{combined\xspace}
\newcommand{\trad}{traditional link prediction\xspace}
\newcommand{\cmark}{\text{\ding{51}}}%
\newcommand{\xmark}{\text{\ding{55}}}%
\newcommand{\edgerand}{EdgeRand\xspace}
\newcommand{\lapgraph}{LapGraph\xspace}
\newcommand{\mes}{\mathcal{M}}
\newcommand{\agg}{\mathcal{A}}
\newcommand{\up}{\mathcal{U}}

\begin{document}

\title{\bf Link Stealing Attacks Against Inductive Graph Neural Networks}

\author{
\rm Yixin Wu\textsuperscript{1}\ \
Xinlei He\textsuperscript{2}\ \
Pascal Berrang\textsuperscript{3}\ \
Mathias Humbert\textsuperscript{4}\ \
\\
Michael Backes\textsuperscript{1}\ \
Neil Zhenqiang Gong\textsuperscript{5}\ \
Yang Zhang\textsuperscript{1}
\\
\\
\textsuperscript{1}\textit{CISPA Helmholtz Center for Information Security}\ \ \ 
\\
\textsuperscript{2}\textit{Hong Kong University of Science and Technology}\ \ \
\textsuperscript{3}\textit{University of Birmingham}\ \ \ 
\\
\textsuperscript{4}\textit{University of Lausanne}\ \ \ 
\textsuperscript{5}\textit{Duke University}\ \ \
}

\date{}

\maketitle

\begin{abstract}

A graph neural network (GNN) is a type of neural network that is specifically designed to process graph-structured data.
Typically, GNNs can be implemented in two settings, including the transductive setting and the inductive setting.
In the transductive setting, the trained model can only predict the labels of nodes that were observed at the training time.
In the inductive setting, the trained model can be generalized to new nodes/graphs.
Due to its flexibility, the inductive setting is the most popular GNN setting at the moment.
Previous work has shown that transductive GNNs are vulnerable to a series of privacy attacks.
However, a comprehensive privacy analysis of inductive GNN models is still missing.
This paper fills the gap by conducting a systematic privacy analysis of inductive GNNs through the lens of link stealing attacks, one of the most popular attacks that are specifically designed for GNNs.
We propose two types of link stealing attacks, i.e., posterior-only attacks and combined attacks.
We define threat models of the posterior-only attacks with respect to node topology and the combined attacks by considering combinations of posteriors, node attributes, and graph features.
Extensive evaluation on six real-world datasets demonstrates that inductive GNNs leak rich information that enables link stealing attacks with advantageous properties.
Even attacks with no knowledge about graph structures can be effective.
We also show that our attacks are robust to different node similarities and different graph features.
As a counterpart, we investigate two possible defenses and discover they are ineffective against our attacks, which calls for more effective defenses.

\end{abstract}

\section{Introduction}

Many types of data can be represented in graphs, such as social networks, molecules, transportation, etc.
Being modeled in a non-Euclidean space, the graph-structured data has imposed formidable challenges on traditional deep learning algorithms.
In order to make full use of the graph-structured data, a graph version of neural networks has emerged~\cite{HYL17, KW17, VCCRLB18, XHLJ19}, which is known as graph neural networks (GNNs).
In practice, GNNs have enabled several successful applications like drug discovery and online service recommendation.
The core idea of GNNs is to leverage neighbor information among nodes to obtain node embeddings.
These node embeddings can then be directly used to predict unknown node labels.
There are mainly two settings for GNNs: the transductive setting and the inductive setting.
In the transductive setting, all the data, including the training dataset and the testing dataset without labels, are used during training.
The GNN model learns from seen data and predicts labels of the testing dataset.
Since the data in the testing dataset is used in the training phase, transductive GNNs cannot be generalized to unseen nodes.
However, in practice, there are a considerable amount of application scenarios where graphs dynamically evolve, e.g., the growth of social networks, thereby making the transductive GNN models less practical.
Unlike GNN models in the transductive setting, inductive GNN models can predict labels of unseen nodes.
Therefore, inductive GNNs are the most popular GNNs at the moment.

Graph-structured data often needs to remain private because data owners spend a large number of resources collecting it or because the data contains inherently sensitive information~\cite{CHGL18, GL16}.
For example, social relationships (encoded by edges in the graph) often represent privacy-sensitive information among users~\cite{WZVMMR18}.
In another example, discovering drugs whose compounds' chemical structures are represented as graphs is extremely costly, making these chemical graph structures valuable assets~\cite{NKB20}.

Previous research has shown that GNN models are vulnerable to privacy attacks, e.g., to membership inference attacks that aim to infer whether a given node~\cite{HWWBSZ21, ONK21} or graph~\cite{WYPY21} is used to train the target GNN model.
More importantly, given that edge information is very privacy-sensitive and is the most significant factor differentiating GNNs from other machine learning models, previous studies have focused on the problem of edge privacy.
He et al.~\cite{HJBGZ21} propose link stealing attacks where different attack strategies are evaluated under eight threat models.
However, they only concentrate on the transductive setting, a less realistic scenario.

In this work, we explore the possibility of link stealing attacks in the inductive setting.
Unlike the transductive setting, where the adversary only needs to feed the nodes' IDs to obtain predictions based on full neighbor information, the inductive setting poses a different challenge.
Here, the adversary relies on their own, often limited and incomplete, neighbor information to construct nodes' subgraphs before feeding them into the target model for prediction.
Although the inductive setting makes the link stealing attacks more challenging, our evaluation shows that our attacks work well with limited or even no neighbor information.
Wu et al.~\cite{WLZL22} propose a query-based attack strategy, LinkTeller, aiming to recover links among nodes of interest in the inductive setting.
LinkTeller has two strong assumptions that are key to the success of the attack: (1) there is a model owner who is aware of the complete graph structure among nodes of interest and can obtain the complete neighbor information each time receiving a query; (2) the adversary is allowed to query the target model with the same set of nodes multiple times.
In our work, we relax these assumptions by (1) only querying the target model with the adversary's own constructed subgraphs that have incomplete or even no neighbor information each time and (2) querying the target model once per node.
Our work stands in a more common scenario where the adversary needs to design attack input, such as nodes' subgraphs, based on their background knowledge.

We propose two types of link stealing attacks that infer whether there exists a link between a given node pair in the target training graph.
We refer to the graph used in the target GNN model training dataset as the target training graph.
The first attack is the \post attack, which relies on the posteriors obtained from a target black-box GNN model to design the attack input features.
The threat model of the \post attack distinguishes between different types of node topology.
As GNNs commonly use two layers (which correspond to the node neighborhood), we consider three types of node topology for our \post attacks, i.e., 0-hop, 1-hop, and 2-hop.
In 0-hop, the adversary has no knowledge about the training graph structure.
In 1-hop/2-hop, the adversary knows 1-hop/2-hop subgraphs for the given node pairs.
Note that the subgraph information is incomplete, as we do not assume the adversary has knowledge about the edge they want to infer.
Additionally, the adversary can have background knowledge, including a shadow dataset containing a graph with node attributes and labels.

We then propose the \comb attack, which combines features used in the traditional link prediction methods with the posteriors to design attack input features.
Previous work solves the link prediction problems using node attributes and graph features that are proven effective in evaluation~\cite{GTMHSSSS14, LK07, BHPZ17}.
Therefore, we define the threat model of the \comb attack against GNNs along three dimensions: in addition to the posteriors, we also consider node attributes and graph features.
For the graph features, we consider network proximity measures based on node neighborhoods~\cite{LK07}.
For the posteriors, we again distinguish between the three types of node topology.
Even if the adversary does not have access to the subgraphs, they can still obtain the 0-hop posteriors.
The adversary can combine posteriors with either node attributes, graph features, or both.
In total, we design seven \comb attacks in which the adversary uses different combinations of these three features (posteriors, node attributes, and graph features).

We evaluate our three \post attacks and seven \comb attacks on six real-world datasets.
Traditional link prediction methods that predict links between nodes in the target training graph are regarded as baseline attacks.
First, the results show that when the adversary has no knowledge about the graph structures, our attacks can achieve outstanding performance, e.g., outperforming the baselines by 14.2\% AUC on DBLP.
This demonstrates that posteriors leak rich information, even with no knowledge about the graph structures, which facilitates link stealing from the inductive GNN models.
Second, our \comb attacks can achieve higher AUC scores than baselines on all datasets, which indicates the inductive GNN models leak extra information to enhance the link stealing attack.
Our attacks outperform the baselines by an average of 6.5\% AUC and up to 12.5\% on DBLP.
Third, we further find that our attacks are more robust to different graph features and different node attributes' cosine similarities (abbreviated as node similarities) than the \trad methods.
This favorable property can help the adversary detect ``surprising'' links that the baseline attacks cannot detect.
It demonstrates that posteriors can enable link stealing attacks with high robustness against different graph features and different node similarities and improve the robustness of traditional link prediction methods.
In addition, experiments show that even if we relax the assumptions, such as using different distributions or different sizes of shadow datasets, different architectures of shadow models, and different architectures of attack models, the link stealing attacks are still effective.
Given these advantageous properties, our attacks represent a severe threat against inductive GNN models in real-world scenarios.
In order to mitigate these attacks, we further investigate a label-only defense mechanism and the state-of-the-art DP-GCN mechanisms~\cite{WLZL22}; the results indicate these defenses are ineffective against our attacks.

In a nutshell, we summarize the contributions as follows:
\begin{itemize}
\item We propose two types of link stealing attacks, i.e., \post attacks and \comb attacks against inductive GNNs.
\item We define the threat models for the \post attacks based on node topology and \comb attacks along three dimensions, i.e., posteriors, node attributes, and graph features.
In total, we propose ten link stealing attacks depending on the adversarial settings.
\item We extensively evaluate our ten attacks on six real-world datasets.
The results show that inductive GNN models leak rich information to enable link stealing attacks that are robust to different graph features and different node similarities.
Moreover, our attacks are still effective in most cases after applying two well-established defense mechanisms.
\end{itemize}

\section{Preliminaries}
\label{section:preliminary}

\subsection{Graph Neural Networks}
\label{section:gnn}

In this paper, we focus on the inductive GNN models trained for node classification tasks that aim to determine nodes' labels.
Given a node $\TargetNode$, the GNN model first learns its node embedding and then leverages the embedding to classify the label of this node.
To generate node embeddings, the GNN model aggregates information from local network neighborhoods using neural networks.
Generally, every node in a $\Radius$-layer GNN has a receptive field of a $\Radius$-hop neighborhood that determines the embedding of the node.
$\Neighbors$ represents a node $\TargetNode$'s neighbors that are $\Radius$-hop away.
The $\Radius$-hop subgraph of node $\TargetNode$ is denoted as $\Subgraph^{\Radius}(\TargetNode)$.
Same as traditional neural networks, GNNs can be of arbitrary depth.
However, as the number of GNN layers increases, the shared neighbors between nodes also quickly grow, and GNNs might suffer from the over-smoothing problem~\cite{CLLLZS20} when $\Radius$ is large.
Thus, $\Radius$ is usually set to two at most.

Given a graph $\Graph =(\NodeSet, \EdgeSet)$ with input node attributes $\{x_{v},\forall v \in \NodeSet \}$, $\NodeSet$ represents the set of all nodes in graph $\Graph$, and $\EdgeSet$ denotes the set of all edges in the graph.
To obtain final node embeddings of $\TargetNode$, node $\TargetNode$ first computes ``transformed'' message of $v$'s k-hop neighborhood $N^{k}(v)$.
Then, node $\TargetNode$ aggregates these ``transformed'' messages from k-hop neighborhood $N^{k}(v)$.
After that, node $\TargetNode$ conducts a non-linear transformation to update its representation.
Formally, each layer of the GNN model can be defined as:
\begin{equation}
\begin{array}{l}
m_{u}^{(l)} = \mes^{(l)}(h_{u}^{(l-1)}, u \in \{N(\TargetNode) \cup \TargetNode\}), \\
z_{\TargetNode}^{(l)} = \agg^{(l)}(\{m^{(l)}_{u}, u \in N(\TargetNode) \}, m_{v}^{(l)}), \\
h_{\TargetNode}^{(l)} = \up(z_{v}^{(l)}),
\end{array}
\end{equation}
where $N(\TargetNode)$ is the neighborhood of $\TargetNode$, $m_{u}^{(l)}$ is the message of node $u$, $z_{\TargetNode}^{(l)}$ denotes the hidden state of $\TargetNode$ in layer $l$, and $h_{\TargetNode}^{(l)}$ represents the representation of node $\TargetNode$ in layer $l$.
The GNN model first initiates all the node embeddings $h_{u}^{(0)}$ using node attributes $x_{u}$.
The $\mes(\cdot)$ function computes messages for nodes including node $\TargetNode$ and its neighbors.
Usually, a linear layer is used as the $\mes(\cdot)$ function structure.
The $\agg(\cdot)$ function then aggregates messages from neighbors.
To prevent information of node $\TargetNode$ itself from getting lost, the $\agg(\cdot)$ function also aggregates the message from node $\TargetNode$ itself.
The $\up(\cdot)$ function usually conducts a non-linear operation to update the representation of node $\TargetNode$.

In this paper, we focus on four different GNN architectures, i.e., GraphSAGE~\cite{HYL17}, Graph Convolutional Network (GCN)~\cite{KW17},  Graph Attention Network (GAT)~\cite{VCCRLB18}, and Graph Isomorphism Network (GIN)~\cite{XHLJ19}.
More details on these models can be found in~\autoref{appendix:gnn_models}.

\subsection{GNN Settings}
\label{section:gnn_setting}

\begin{figure}[!t]
\centering
\includegraphics[width=\columnwidth]{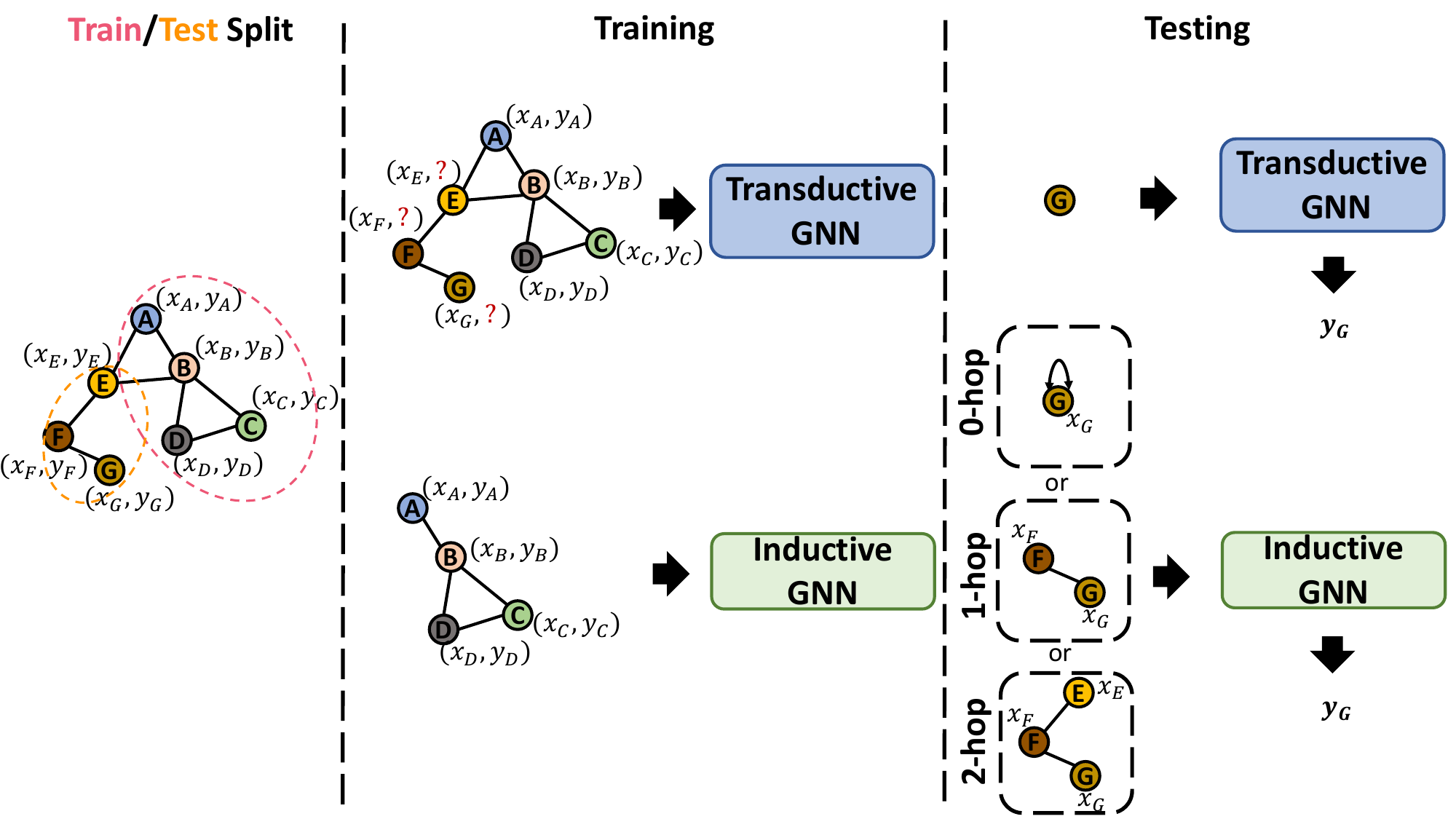}
\caption{Comparison between the transductive and the inductive settings.}
\label{figure:gnn_setting}
\end{figure}

As illustrated in~\autoref{figure:gnn_setting}, there are two settings for GNNs: transductive setting and inductive setting.
In the transductive setting, the GNN model learns the embeddings of nodes that are contained in the training graph $G^{\textit{Train}}$.
Note that in the transductive setting, the entire graph $G$ with its node attributes $X_{V}$ can be observed in both the training and testing phases, i.e., $G=G^{\textit{Train}}=G^{\textit{Test}}$.
During the training phase, we use $G$ with node attributes and only a small subset of labels.
During the testing phase, the users feed the identifiers of unlabeled nodes in $G$ to query the GNN model and obtain prediction results.
It means that the users including the adversary can obtain the posteriors based on full neighbor information, even though they do not have any knowledge of the neighbor information or node attributes at the testing time.

When there is a need for predicting unseen nodes, the inductive setting comes on stage.
In the inductive setting, the GNN model learns aggregation functions that can induce the embeddings of unseen nodes.
As illustrated in~\autoref{figure:gnn_setting}, the training and testing datasets have independent graphs.
The model owner first trains the GNN model using the training graph $G^{\textit{Train}}$ of the training dataset $D^{\textit{Train}}$.
Then, at the testing time, given a target node $\TargetNode$ in the testing graph $G^{\textit{Test}}$, the users need to construct a $k$-hop subgraph $g^{k}_{(v)}$ containing node attributes and graph structure of $N(v)$ based on their knowledge and then feed it into the trained GNN model to predict the label of $\TargetNode$.
We refer to the query using $k$-hop subgraph $g^{k}_{(v)}$ as $k$-hop query, where $k \in \{0,1,2\}$.
The classification ability can be generalized to previously unseen nodes, thus the inductive GNNs have broader application scenarios.

He et al.~\cite{HJBGZ21} have demonstrated that transductive GNNs are vulnerable to link stealing attacks.
Here, the posteriors are predicted based on the full neighbor information including the link the adversary aims to infer.
However, in the inductive setting, since the trained models have not seen the neighbor information of the testing data during the training phase, the user (including the adversary) can only obtain posteriors based on the subgraphs they constructed.
Consequently, the neighbor information is always incomplete because of the link the adversary aims to infer, making the link stealing attacks against inductive GNNs significantly more challenging than those against transductive GNNs.

\begin{table*}[!t]
\caption{Attack input features for all attacks.}
\label{table:attack_feature}
\centering
\renewcommand{\arraystretch}{1.1}
\scalebox{0.85}{
\begin{tabular}{c|c|c c c|c|c|c|c|c c c|c|c}
\toprule
Attack & Attack & \multicolumn{3}{c|}{Posteriors} & Node & Graph & Attack & Attack & \multicolumn{3}{c|}{Posteriors} & Node &  Graph \\
Method & Model  & 0-hop & 1-hop & 2-hop & Attribute &  Feature & Method & Model  & 0-hop & 1-hop & 2-hop & Attribute & Feature \\
\midrule
Attack-0 & $A_{(p, ~\cdot, ~\cdot)}^{0}$ & $\cmark$ & $\xmark$ & $\xmark$ & $\xmark$ & $\xmark$ & Attack-7 & $A_{(p, ~\cdot, ~g)}^{2}$ & $\xmark$ & $\xmark$ & $\cmark$ & $\xmark$ & $\cmark$  \\
Attack-1 & $A_{(p, ~\cdot, ~\cdot)}^{1}$ & $\xmark$ & $\cmark$ & $\xmark$ & $\xmark$ & $\xmark$ & Attack-8 & $A_{(p, ~n, ~g)}^{1}$ & $\xmark$ & $\cmark$ & $\xmark$ & $\cmark$ & $\cmark$\\
Attack-2 & $A_{(p, ~\cdot, ~\cdot)}^{2}$ & $\xmark$ & $\xmark$ & $\cmark$ & $\xmark$ & $\xmark$ & Attack-9 & $A_{(p, ~n, ~g)}^{2}$ & $\xmark$ & $\xmark$ & $\cmark$ & $\cmark$ & $\cmark$\\
Attack-3 & $A_{(p, ~n, ~\cdot)}^{0}$ & $\cmark$ & $\xmark$ & $\xmark$ & $\cmark$ & $\xmark$ & Baseline-0 & $B_{(\cdot, ~n, ~\cdot)}$ & $\xmark$ & $\xmark$ & $\xmark$ & $\cmark$ & $\xmark$\\
Attack-4 & $A_{(p, ~n, ~\cdot)}^{1}$ & $\xmark$ & $\cmark$ & $\xmark$ & $\cmark$ & $\xmark$ & Baseline-1 & $B_{(\cdot, ~\cdot, ~g)}$ & $\xmark$ & $\xmark$ & $\xmark$ & $\xmark$ & $\cmark$\\
Attack-5 & $A_{(p, ~n, ~\cdot)}^{2}$ & $\xmark$ & $\xmark$ & $\cmark$ & $\cmark$ & $\xmark$ & Baseline-2 & $B_{(\cdot, ~n, ~g)}$ & $\xmark$ & $\xmark$ & $\xmark$ & $\cmark$ & $\cmark$\\
Attack-6 & $A_{(p, ~\cdot, ~g)}^{1}$ & $\xmark$ & $\cmark$ & $\xmark$ & $\xmark$ & $\cmark$ & & & & & & \\
\bottomrule
\end{tabular}
}
\end{table*}

\section{Our Attacks}

\subsection{Problem Definition}

Suppose there is an inductive target GNN model $M_{T}$ trained on the target training dataset $D_{\textit{Target}}^{\textit{Train}}$.
The model owner provides users with an inference API to obtain prediction results, which is often the case in practice.
Like general users, the adversary can query $M_{T}$, but they utilize these prediction results for malicious purposes.
Specifically, given a node pair $(u, v)$ in the target training graph $G^{\textit{Train}}_{\textit{Target}}$, the adversary's goal is to infer whether there exists a link between nodes $u$ and $v$.

\subsection{Threat Model}

\mypara{Adversary's Background Knowledge}
As mentioned above, we first assume that the adversary can only have black-box access to the target GNN model $M_{T}$, which is the most challenging setting~\cite{SSSS17, SBBFZ20, SZHBFB19}.
To obtain the posteriors of $\TargetNode$, the adversary needs to construct $g^{k}_{v}$ that contains node attributes and graph structures.
Node attributes can be easily acquired.
For instance, the adversary can crawl social media content to extract node attributes for users in a social network.
We discuss graph structures with different node topologies based on the adversary's knowledge in the following subsection.
We also assume that the adversary can have a shadow dataset $D_{\textit{Shadow}}$ that contains node attributes, graphs, and ground-truth labels.
Following previous work~\cite{SSSS17}, we assume we can get $D_{\textit{Shadow}}$ that comes from the same distribution as $D_{\textit{Target}}$.
Note that $D_{\textit{Target}}$ and $D_{\textit{Shadow}}$ do not have any overlap in terms of nodes or edges.
The adversary can train a shadow model $M_{S}$ to mimic the behavior of $M_{T}$ using $D_{\textit{Shadow}}$.
The later evaluation results show that the distribution/size of the shadow dataset and the architecture of the shadow model only has a slight effect on the attack performance.

\mypara{Posterior-Only Attacks}
This type of attack only uses the posteriors of the node pair $(u, v)$ obtained from the GNN models to design attack input features.
Different from the previous work~\cite{HJBGZ21, WLZL22} where the target GNN models only take nodes of interest as input and leverage full neighbor information, we consider the most common scenario where the adversary needs to design attack input, i.e., nodes' subgraphs, based on their background knowledge.
As the GNN models are in the inductive setting, we define the threat model of \post attacks by categorizing the node topology in 0-hop, 1-hop, and 2-hop.
We consider these three cases because there are usually at most two layers in inductive GNN models (see~\autoref{section:gnn}).
In the 0-hop case, the adversary only knows the node attributes of $u$ and $v$.
In other words, the adversary has no neighbor information.
As the input format of the GNN model is a graph, the adversary can add a self-loop edge for node $u$ and $v$ respectively to construct graphs and conduct a 0-hop query to get the node posteriors.
In the 1-hop case, the adversary has knowledge about 1-hop subgraphs for $u$ and $v$ except for the link aiming to infer, meaning that the 1-hop subgraph is actually incomplete.
The adversary can feed the incomplete 1-hop subgraphs into the target GNN model and perform a 1-hop query to get node posteriors for two nodes.
The 2-hop case is similar to 1-hop except the adversary knows the incomplete 2-hop subgraphs of $u$ and $v$ (the link between $u$ and $v$ is missing).

\mypara{Combined Attacks}
Inspired by previous work~\cite{LK07, GTMHSSSS14}, we propose \comb attacks that combine posteriors with traditional link prediction features, i.e., node attributes and graph features.
We define threat models of \comb attacks against GNNs along three dimensions, i.e., posteriors, node attributes, and graph features.
Specifically, we consider three cases that combine posteriors with either of the other two features or both of them.
In the first case, besides obtaining posteriors, the adversary can also leverage node attributes, as neighbor node attributes have an intrinsic similarity.
Thus, the adversary can combine posteriors with node attributes to design attack input features.
Also, the posteriors can be further divided into three types, i.e., 0-hop, 1-hop, and 2-hop.
In the second case, the adversary can perform the 1-hop/2-hop query using 1-hop/2-hop subgraphs.
They can further use subgraphs to generate graph features, e.g., common neighbors, which can reflect the connectivity between node neighborhoods.
The natural intuition is that node pairs with higher connectivity are more likely to be linked in a graph structure.
In the last case, the adversary makes full use of all background knowledge, i.e., knowledge of all three dimensions, to launch the link stealing attacks.
Note that, in the latter two cases, the adversary can only generate graph features when performing 1-hop/2-hop queries, as they do not have any graph information in the 0-hop query case.

\subsection{Attack Methodology}

\mypara{Shadow Model Training}
The adversary first divides $D_{\textit{Shadow}}$ evenly into two disjoint splits.
As it is an inductive setting, each split contains an independent graph with node attributes, edges, and labels.
The first split is treated as the shadow training dataset $D^{\textit{Train}}_{\textit{Shadow}}$ to train $M_{S}$, while the second half is considered as the shadow testing dataset $D^{\textit{Test}}_{\textit{Shadow}}$ to evaluate $M_{S}$.

\mypara{Attack Model Training}
The attack model is a binary classifier that can predict if two given nodes have a connection in $G_{\textit{Target}}^{\textit{Train}}$.
Therefore, the output of the attack model is 1/0, indicating whether there exists a link between the given pair or not.
Its input is derived from the GNN's outputs (i.e., posteriors), node attributes, and network proximity measures based on node neighborhoods.
To construct the attack training dataset $D_{\textit{Attack}}^{\textit{Train}}$, the adversary first needs to query $M_{S}$ using subgraphs with node attributes from the shadow training dataset to get the posteriors.
The attack input features of the \post attack are only designed over posteriors.
The adversary can conduct the 0-hop query, 1-hop query, or 2-hop query based on the background knowledge, which leads to different attack taxonomies of the \post attacks.
Depending on the background knowledge, the adversary can also leverage either node attributes or graph features, or both of them.
Different types of features and query methods used to design attack input features together result in different taxonomies of the \comb attacks.
The attack models of different attacks are referred to as $A_{\alpha}^{\beta}$, where $A$ is short for the attack, $\beta$ is the number of hops queried by the adversary, and $\alpha = (p,~n,~g)$ is a combination of characters: $p$, $n$, and $g$ represent posteriors, node attributes, and graph features, respectively
Attack input features for all attacks are shown in~\autoref{table:attack_feature}.

\begin{table}[!t]
\caption{Pairwise operations.}
\label{table:pairwise}
\centering
\renewcommand{\arraystretch}{1.1}
\scalebox{0.85}{
\begin{tabular}{c|c}
\toprule
Operator & Definition\\
\midrule
$\textit{Hadamard}$ & $f_{i}(u) * f_{i}(v)$ \\
$\textit{Average}$ & $\frac{f_{i}(u) + f_{i}(v)}{2}$ \\
$\textit{Weighted-L1}$ & $|f_{i}(u) - f_{i}(v)|$ \\
$\textit{Weighted-L2}$ & $|f_{i}(u) - f_{i}(v)|^2$\\
\bottomrule
\end{tabular}
}
\end{table}

\begin{figure}[!t]
\centering
\includegraphics[width=\columnwidth]{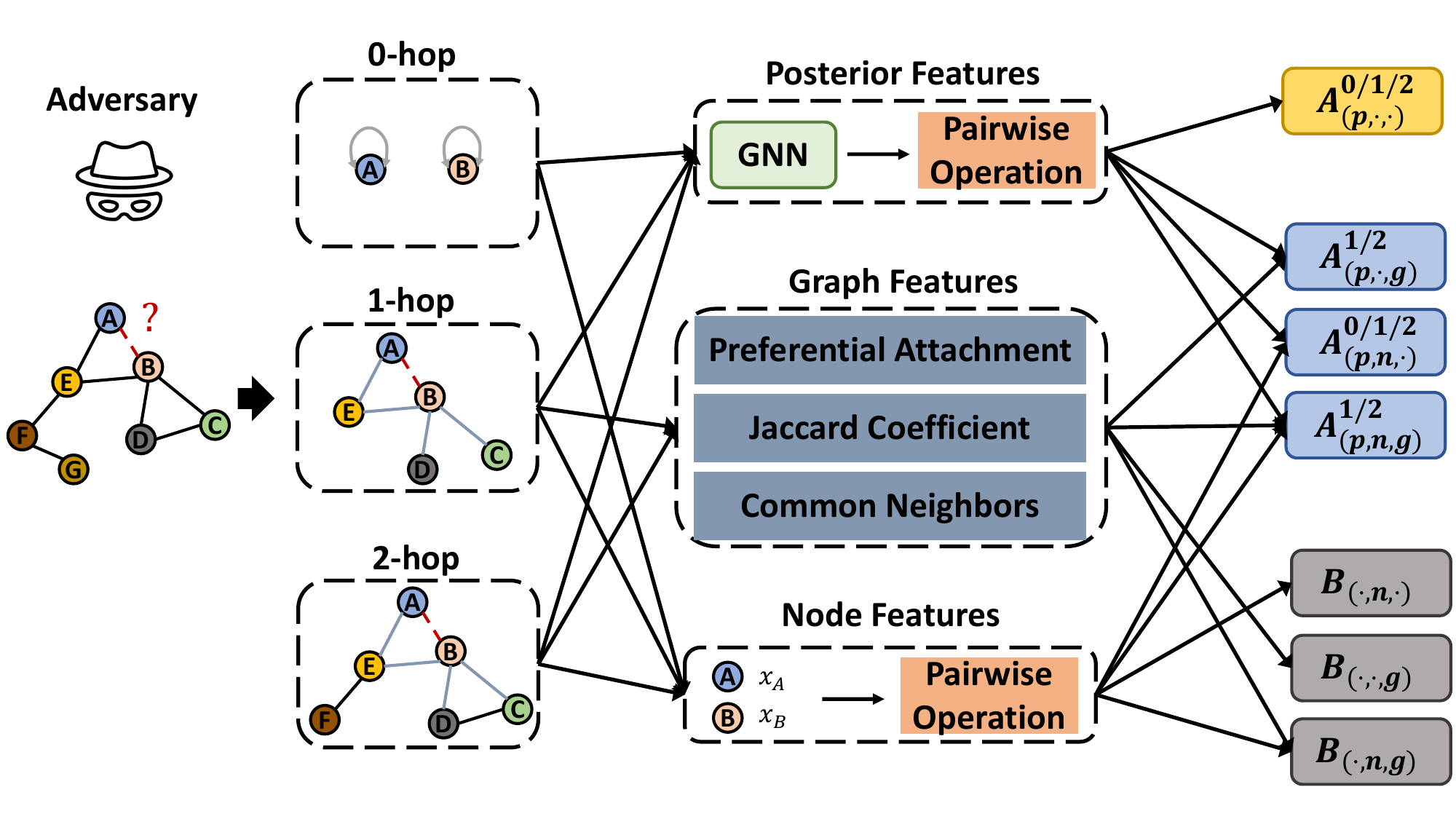}
\caption{Overview of the proposed \post attack, \comb attack and baselines.}
\label{figure:overview}
\end{figure}

\mypara{Taxonomy of Posterior-Only Attacks (Attack-0 to Attack-2)}
We present an overview of the proposed posterior-only attacks, combined attacks, and baselines in~\autoref{figure:overview}.
The \post attacks leverage posteriors to design attack input features.
For a given node pair, the adversary first feeds each node with its subgraph and node attributes into the GNN model to get the posteriors.
As the attack input features should be the same regardless of the input order of nodes, we take a step further to process the posteriors of the given node pair.
Specifically, we follow the strategy in~\cite{HJBGZ21} and leverage four pairwise, commutative operations to ensure attack input features are the same even if nodes are presented in a different order for the given pair.
These pairwise operations are summarized in \autoref{table:pairwise}.
In later~\autoref{section:evaluation}, we perform a grid search to determine the exact pairwise operation for the proposed attacks.
After feeding posteriors into these operations, we concatenate these results as the attack input feature for each pair.
For \post attacks (Attack-0/1/2), the attack models are defined as $A_{(p, ~\cdot, ~\cdot)}^{\beta}$.
To launch these attacks, we adopt MLP models that take features generated from 0/1/2-hop posteriors as input.

\mypara{Taxonomy of Combined Attacks (Attack-3 to Attack-9)} 
In addition to using posteriors, the \comb attack combines posteriors with either node attributes or graph features or both to design attack input features.

For the \comb attacks that use posteriors and node attributes (Attack-3/4/5), the attack model is defined as $A_{(p, ~n, ~\cdot)}^{\beta}, \beta \in \{0, 1, 2 \}$, as we assume the adversary has the knowledge to perform the 0/1/2-hop queries in this attack.
Node attributes for node pairs also face the node order issue.
Therefore, we apply the pairwise operations in~\autoref{table:pairwise} for node attributes as well.
Due to its high dimensionality, we only apply the Hadamard operation for node attributes.
The attack model $A_{(p, ~n, ~\cdot)}^{\beta}$ is a multiple-input MLP model that takes features generated by 0/1/2-hop posteriors and node attributes separately as input.

For the \comb attacks that use posteriors and graph features (Attack-6/7), the attack model is defined as $A_{(p, ~\cdot, ~g)}^{\beta}, \beta \in \{1, 2\}$, as we assume the adversary has the knowledge to perform the 1-hop/2-hop queries in this attack.
Note that there is no combination of 0-hop posteriors and graph features, as we assume the adversary has no knowledge about graph structures when they perform the 0-hop query.
Following previous work~\cite{LK07}, we focus on network proximity measures, including common neighbors, Jaccard coefficient, and preferential attachment, and concatenate these three features to be our graph features.
The attack model $A_{(p, ~\cdot, ~g)}^{\beta}$ is a multiple-input MLP model that takes features generated from 1-hop/2-hop posteriors and graph features separately as input.

For the \comb attacks that use posteriors, node attributes, and graph features (Attack-8/9), the attack model is defined as $A_{(p, ~n, ~g)}^{\beta}, \beta \in \{1, 2\}$.
The attack model $A_{(p, ~n, ~g)}^{\beta}$ is a multiple-input MLP model that takes features generated from 1-hop/2-hop posteriors, node attributes, and graph features separately as input.

\mypara{Baselines~\cite{GTMHSSSS14, LK07, BHPZ17}}
We compare the \post attacks and \comb attacks with three baselines.
These baselines are used to perform traditional link prediction tasks that aim to infer whether there exists a link between a given node pair in the graph of interest.
We can exploit these traditional link prediction methods to perform link stealing attacks, as the graph of interest is set to $G_{\textit{Target}}^{\textit{Train}}$.
Baseline-0 only uses node attributes, and the attack model is referred to as $B_{(\cdot, ~n, ~\cdot)}$.
Baseline-1 only uses graph features, and the attack model is referred to as $B_{(\cdot, ~\cdot, ~g)}$.
Baseline-2 uses node attributes and graph features, and the attack model is referred to as $B_{(\cdot, ~n, ~g)}$.
The attack model $B_{(\cdot, ~n, \cdot)}$ and $B_{(\cdot, \cdot, ~g)}$ are MLP models that take features generated from either node attributes or graph features as input.
The attack model $B_{(\cdot, ~n, ~g)}$ is a multiple-input MLP model that takes features generated from node attributes and graph features separately as input.

\mypara{Link Stealing}
To enable the link stealing attack for a given node pair, the adversary first performs a 0-hop, 1-hop, or 2-hop query to obtain posteriors based on the background knowledge of node topology.
Then, the adversary leverages posteriors alone or combines them with node attributes and graph features to generate attack input features.
Finally, these input features are fed into the attack model to determine whether the given node pair has a link in $G_{\textit{Target}}^{\textit{Train}}$.

\section{Evaluation}
\label{section:evaluation}

In this section, we first introduce the details of the experimental setup.
Second, we present the target performance and the attack performance and investigate to what extent our attacks can be a practical threat against inductive GNN models via relaxing assumptions.
Third, we investigate the robustness of our attacks against different graph features and different node similarities.
At last, we evaluate the performance of two defense mechanisms.

\subsection{Experimental Setup}
\label{section:setup}

\begin{table}[!t]
\caption{Dataset statistics.}
\label{table:dataset_statistics}
\centering
\renewcommand{\arraystretch}{1.1}
\scalebox{0.85}{
\begin{tabular}{l|c c c c}
\toprule
Dataset  & \# Nodes & \# Edges & Density & \# Classes \\
\midrule
Cora & 2,995 & 8,416 & 0.00215 & 7 \\
Pubmed & 19,717 & 44,324 & 0.00028 & 3\\
DBLP & 17,716 & 52,867 & 0.00039 & 4  \\
Photo & 7,650 & 143,663 & 0.00420 & 8 \\
CS & 13,752 & 287,209 & 0.00267 & 10 \\
LastFM & 7,624 & 63,236 & 0.00109 & 18 \\
\bottomrule
\end{tabular}
}
\end{table}

\mypara{Datasets}
We evaluate our attacks on six public datasets including Cora-ML (abbreviated as Cora)~\cite{BG182}, Pubmed~\cite{SNBGGE08}, DBLP~\cite{PWZZW16}, the AmazonCoBuy dataset for photos (abbreviated as Photo), the AmazonCoBuy dataset for computers (abbreviated as CS)~\cite{MTSH15}, and LastFM Asia Social Network (abbreviated as LastFM)~\cite{RS20}.
Among them, Cora, Pubmed, and DBLP are citation networks where nodes are publications and edges are citation links among these publications.
The two Amazon datasets are part of the Amazon co-purchase graph where nodes are items, and edges between two items indicate that they have been purchased together.
The LastFM dataset contains a social network where nodes are LastFM users from Asian countries, and edges are mutual follower relationships between them.
General statistics, e.g., the number of edges and density, for all datasets are summarized in~\autoref{table:dataset_statistics}.
Among all datasets, Pubmed and DBLP are two datasets with sparser density.

\mypara{Dataset Configurations} 
We first randomly split the whole graph in half according to the number of nodes for each dataset.
The first half is treated as the target dataset $D_{\textit{Target}}$, while the second half constructs the shadow dataset $D_{\textit{Shadow}}$.
These two parts are disjoint.
We discard the edges between nodes that end up in different splits; this affects the degree distribution of the resulting graphs at a minimal level.
We further divide $D_{\textit{Target}}$ into the target training dataset $D_{\textit{Target}}^{\textit{Train}}$ and target testing dataset $D_{\textit{Target}}^{\textit{Test}}$ in an 8:2 ratio.
$D_{\textit{Target}}^{\textit{Train}}$ is used to train the target model $M_{T}$, while $D_{\textit{Target}}^{\textit{Test}}$ can evaluate $M_{T}$'s performance on the original task.
The same processing procedure is applied to the shadow dataset $D_{\textit{Shadow}}$.
Note that each dataset contains an independent graph, node attributes, and node labels.
For instance, $G_{\textit{Target}}^{\textit{Train}}$, along with its node attributes and node labels, is included in $D_{\textit{Target}}^{\textit{Train}}$.
We use $D_{\textit{Shadow}}$ to generate the attack training dataset $D_{\textit{Attack}}^{\textit{Train}}$, while $D_{\textit{Target}}$ is leveraged to construct the attack testing dataset $D_{\textit{Attack}}^{\textit{Test}}$.
The attack dataset contains input feature vectors of node pairs and labels indicating whether two nodes are linked or not.
We refer to node pairs that are linked as positive pairs and node pairs that are not linked as negative pairs.
To build $D_{\textit{Attack}}^{\textit{Train}}$, we first select all node pairs that are actually connected in the shadow training graph $G_{\textit{Shadow}}^{\textit{Train}}$ as positive pairs.
Following the negative sampling approach in previous work~\cite{HJBGZ21}, we then randomly sample the same number of node pairs that are not connected in $G_{\textit{Shadow}}^{\textit{Train}}$ as negative pairs.
We apply the same processing procedure on the $D_{\textit{Target}}^{\textit{Train}}$ to construct the $D_{\textit{Attack}}^{\textit{Test}}$ that can evaluate all of our attacks.

\mypara{Metrics}
Following previous work~\cite{BHPZ17, HJBGZ21}, we rely on the AUC score (area under the ROC curve) to evaluate the attack performance of link stealing attacks.
We also leverage accuracy to evaluate the target model's performance.

\noindent \mypara{Implementation Details}
To train the models, we either follow the experimental setup of previous related work~\cite{HJBGZ21,HWWBSZ21,SHHZ22,WLZL22} or conduct a grid search to find the best set of hyper-parameters.
Below, we describe the final hyper-parameter set and architectures of the target, shadow, baseline, and attack models.
We also report the search space of the hyper-parameters in~\autoref{appendix:grid_search}.

\noindent \emph{Target models.}
We leverage four GNN architectures, i.e., GraphSAGE, GCN, GAT, and GIN, to build the target models and shadow models.
We follow previous work~\cite{HJBGZ21,HWWBSZ21,WLZL22} and use a two-layer architecture with a full-neighbor sampler for each target model.
We set the number of neurons to 128 in the hidden layer.
The unit size of the output layer is determined by the number of classes of the original task.
Each layer employs a ReLU activation and 0.5 dropout rate to reduce overfitting.
In addition, the first layer of the GAT model has two attention heads, and the second layer only has one attention head.
All models use cross-entropy as the loss function and Adam as the optimizer.
The initial learning rate is set to 0.001.
Both the target models and shadow models are trained for 200 epochs.

\noindent \emph{Baseline models.}
We use a three-layer MLP for Baseline-0.
The hidden unit size of the first layer and the second layer are set to 128 and 32, respectively.
The number of neurons in the output layer is set to two because the link stealing attack is a binary classification task.
As the dimension of graph features is much smaller than that of node attributes, we only use an MLP model with two layers for Baseline-1.
The hidden unit size is set to 16.
Regarding Baseline-2, the two inputs are first fed into two sub-networks simultaneously.
The sub-network for node attributes consists of three linear layers with 256 neurons, 64 neurons, and 8 neurons, respectively.
Another sub-network is composed of one linear layer with one neuron.
We concatenate the 8-dimensional embedding and 1-dimensional embedding together and feed them into a linear layer for link stealing.
These baseline models also have the same activation function and drop rate in each layer.
In addition, the loss function, optimizer, initial learning rate, and training epochs are the same as the target model.
We use a cosine annealing scheduler to tune the learning rate in the training process.

\noindent \emph{Posterior-only attack models.}
We utilize a three-layer MLP as an attack model for all \post attacks.
These three linear layers have 128, 32, and 2 neurons, respectively.
The loss function, optimizer, initial learning rate with its scheduler, and training epochs are the same as the baseline models.

\noindent \emph{Combined attack models.}
The attack model $A_{(p, ~n, ~\cdot)}^{\beta}$ is composed of two sub-networks and a linear layer.
The first sub-network receives features generated from node attributes and comprises three linear layers with 128 neurons, 64 neurons, and 16 neurons, respectively.
The second sub-network, consisting of two layers with 64 and 16 neurons, receives features derived from posteriors.
These two 16-dimensional embeddings are concatenated and fed into a linear layer to make predictions.
The attack model $A_{(p, ~\cdot, ~g)}^{\beta}$ also has two sub-networks and a linear layer.
However, the first sub-network receives graph features and consists of two linear layers with 16 neurons and 4 neurons, respectively.
The sub-network that receives posteriors is composed of three layers with 128 neurons, 64 neurons, and 16 neurons, respectively.
The 4-dimensional and 16-dimensional embeddings are concatenated and fed into a linear layer to make predictions.
The attack model $A_{(p, ~n, ~g)}^{\beta}$ consists of three sub-networks and a linear layer.
The first sub-network that receives node attributes is composed of three layers with 128 neurons, 64 neurons, and 16 neurons, respectively.
The second sub-network is the same as the first one, but it is used to receive posteriors.
The third sub-network comprises only one linear layer with four neurons and receives graph features.
These three embeddings are concatenated and fed into a linear layer to make predictions.
The loss function, optimizer, initial learning rate with its scheduler, and training epochs of the \comb attacks are the same as the baseline models.
We show that, even if the attack model has different architectures, i.e., different number of layers, we can still get similar attack performance (see \autoref{appendix:ablation_diff_attack_model}).

Our codes\footnote{\url{https://github.com/yxoh/link_steal_pets2024}.} are mainly implemented with PyTorch,\footnote{\url{https://pytorch.org/}.} Networkx,\footnote{\url{https://networkx.org/}.} and DGL.\footnote{\url{https://www.dgl.ai/}.} 
All experiments are carried out on an NVIDIA DGX Server.

\begin{table}[!t]
\caption{Test accuracy of the original tasks for different GNN architectures on six different datasets.}
\label{table:target_performance}
\centering
\renewcommand{\arraystretch}{1.1}
\scalebox{0.85}{
\begin{tabular}{c|c c c c|c}
\toprule
& \multicolumn{4}{c|}{Target Model} & Baseline\\
Dataset  & GraphSAGE & GIN & GAT & GCN & MLP\\
\midrule
Cora & 0.773 & 0.757 & 0.737 & 0.763 & 0.747\\
Pubmed & 0.871 & 0.855 & 0.865 & 0.857 & 0.850\\
DBLP & 0.739 & 0.727 & 0.746 & 0.746  & 0.726\\
Photo & 0.935 & 0.877 & 0.859 & 0.855 & 0.790\\
CS & 0.850 & 0.778 & 0.820 & 0.805 & 0.774\\
LastFM  & 0.752 &  0.738 &  0.721 &  0.760 & 0.718\\
\bottomrule
\end{tabular}
}
\end{table}

\subsection{Target Model Performance}

We first present in \autoref{table:target_performance} the accuracy of the original node classification tasks for four GNN architectures on six different datasets.
We can observe that all GNN models can achieve great target performance on all datasets.
Following previous studies~\cite{HWWBSZ21,WYPY20}, we construct a 2-layer MLP model as the baseline to undertake the same target classification tasks.
We observe that our target GNN models consistently outperform the baseline.
For instance, on the Photo dataset, the baseline MLP achieves an accuracy of 0.790, while GraphSAGE, GIN, GAT, and GCN achieve accuracies of 0.935, 0.877, 0.859, and 0.855, respectively.
This demonstrates that employing graph neural networks to jointly use the node attributes and graph structures can well improve the node classification performance.
We can also observe that the choice of architecture has a more significant impact on dense datasets.
It is easily understood that these datasets have more complex neighbor relationships; thus, different aggregating methods can result in a larger variance in the target model performance.
For instance, we consider the gap between the best and worst target performance on each dataset.
It is 1.6\% on the Pubmed dataset, while 8.0\% on the Photo dataset.
Compared to the Photo dataset, the Pubmed dataset is much sparser.

\subsection{Our Attacks}

\begin{table*}[!t]
\caption{Attack performance on all six datasets.
The average AUC score of five runs is reported.
Both the target model and the shadow model are GraphSAGE.
The best performance of the \post attacks and the \comb attacks are highlighted in bold (A: Attack, B: Baseline).}
\label{table:attack_performance_all}
\centering
\renewcommand{\arraystretch}{1.1}
\scalebox{0.85}{
\begin{tabular}{c|c c c|c c c|c c c c c c c}
\toprule
& \multicolumn{3}{c|}{Baseline~\cite{GTMHSSSS14, LK07, BHPZ17}} & \multicolumn{3}{c}{Posterior-Only Attack} & \multicolumn{7}{|c}{Combined Attack}  \\
Dataset & B0 & B1 & B2  & A0 & A1 & A2 & A3 & A4 & A5 & A6 & A7 & A8 & A9 \\
\midrule
Cora & 0.748 & 0.820 & 0.769 & 0.859 & 0.849 & 0.849 & 0.876 & 0.876 & 0.875 & 0.882 & 0.884 & 0.908 & \textbf{0.909} \\
Pubmed & 0.876 & 0.820 & 0.889 & 0.768 & 0.806 & 0.809 & 0.889 & 0.895 & 0.897 & 0.881 & 0.882 & \textbf{0.939} & \textbf{0.939} \\
DBLP & 0.692 & 0.787 & 0.804 & 0.781 & 0.821 & 0.822 & 0.834 & 0.873 & 0.872 & 0.879 & 0.903 & 0.924 & \textbf{0.929} \\
Photo & 0.813 & 0.930 & 0.878 & 0.877 & 0.898 & 0.898 & 0.892 & 0.916 & 0.915 & 0.967 & \textbf{0.968} & 0.946 & 0.946 \\
CS & 0.821 & 0.932 & 0.863 & 0.817 & 0.838 & 0.845 & 0.869 & 0.890 & 0.893 & 0.955 & \textbf{0.956} & 0.941 & 0.940 \\
LastFM & 0.798 & 0.786 & 0.866 & 0.850 & 0.869 & 0.867 & 0.883 & 0.909 & 0.911 & 0.919 & 0.921 & 0.929 & \textbf{0.930} \\
\bottomrule
\end{tabular}
}
\end{table*}

\begin{table*}[!t]
\caption{Attack performance of \post attacks on six datasets when the target model is GraphSAGE and the shadow model is one of four architectures we mentioned in \autoref{section:preliminary}.
The average AUC score of five runs is reported.
The best performance of the \post attacks is highlighted in bold.}
\label{table:attack_performance_2}
\centering
\renewcommand{\arraystretch}{1.1}
\scalebox{0.85}{
\begin{tabular}{c|c|c c c c|c|c|c c c c}
\toprule
& & \multicolumn{4}{c|}{Architecture} & & & \multicolumn{4}{c}{Architecture}  \\
Dataset & Method & GraphSAGE & GIN & GAT & GCN & Dataset & Method & GraphSAGE & GIN & GAT & GCN \\
\midrule
\multirow{3}{*}{Cora} & Attack-0 & 0.859 & 0.853 & 0.858 & \textbf{0.860} & \multirow{3}{*}{Pubmed} & Attack-0 & \textbf{0.768} & 0.764 & 0.763 & 0.765 \\ & Attack-1 & 0.849 & 0.846 & \textbf{0.850} & 0.842 & & Attack-1 & \textbf{0.806} & 0.805 & 0.800 & 0.801 \\ & Attack-2 & 0.849 & 0.843 & \textbf{0.852} & 0.843 & & Attack-2 & 0.809 & \textbf{0.810} & 0.805 & 0.806 \\
\midrule
\multirow{3}{*}{DBLP} & Attack-0 &   0.781 & 0.779 & \textbf{0.782} & 0.779 & \multirow{3}{*}{Photo} & Attack-0 &  \textbf{0.877} & 0.722 & 0.783 & 0.750 \\ 
& Attack-1 &   0.821 & 0.820 & 0.818 & \textbf{0.824} & & Attack-1 & \textbf{0.898} & 0.785 & 0.801 & 0.810 \\ 
& Attack-2 &  0.822 & 0.824 & 0.821 & \textbf{0.829}  & & Attack-2 &  \textbf{0.898} & 0.786 & 0.821 & 0.828 \\ 
\midrule
\multirow{3}{*}{CS} & Attack-0 &   \textbf{0.817} & 0.770 & 0.798 & 0.785  & \multirow{3}{*}{LastFM} & Attack-0 & \textbf{0.850} & 0.804 & 0.836 & 0.794 \\ 
& Attack-1 &  0.838 & \textbf{0.840} & 0.837 & 0.823 & & Attack-1 & 0.869 & 0.855 & \textbf{0.870} & 0.833 \\ 
& Attack-2 &  \textbf{0.845} & 0.837 & 0.839 & 0.810 & & Attack-2 &  0.867 & 0.867 & \textbf{0.871} & 0.831 \\
\bottomrule
\end{tabular}
}
\end{table*}

We show the link stealing attack performance of the \post attacks, \comb attacks, and baseline attacks in \autoref{table:attack_performance_all}.
Due to the space limitation, we only show the results when both $M_{T}$ and $M_{S}$ are GraphSAGE.
We observe that our attacks with 0-hop posteriors perform well.
When the adversary has no knowledge about the graph structures, they can only perform Baseline-0, Attack-0, and Attack-3.
We find that Attack-0 and Attack-3 outperform Baseline-0 on most of the datasets.
For instance, on DBLP, Attack-3/Attack-0 outperforms Baseline-0 by 14.2\%/8.9\% AUC.
It indicates that the inductive GNN can leak enough information to enable high-performing link stealing attacks, even only feeding into node attributes with self-loop edges.
In other words, our attack is still effective even with no knowledge about the graph structures.

When the adversary has both node attributes and graph structures, they can perform all \post attacks, \comb attacks, and baseline attacks.
Specifically, on the Cora, DBLP, and LastFM datasets, the \post attacks are able to outperform all baseline attacks.
The adversary can rely on the combined attacks to outperform baselines on the other three datasets.
We also find that our attacks using the 1-hop query can achieve a performance similar to the 2-hop query.
Compared to Attack-1, Attack-2 can only achieve an improvement of 0.15\% AUC on average on all datasets.
We can also get the same conclusion on the comparison of other 1-hop attacks and 2-hop attacks, e.g., Attack-8 and Attack-9.
This demonstrates that inductive GNNs leak rich information even though the adversary only has limited knowledge about the graph structures (e.g., 1-hop subgraph).
Overall, our attacks outperform the baselines by an average of 7.23\% AUC on all six datasets, demonstrating that inductive GNNs are indeed more vulnerable to the proposed link stealing attacks than the baseline attacks.
The posteriors are the most useful information, as they enable well-performing attacks and enhance the traditional link prediction methods.
With more information, the \comb attacks always outperform the \post attacks.
However, the attack model $A_{(p, ~\cdot, ~g)}^{*}$ surpasses the $A_{(p, ~n, ~g)}^{*}$ on the Photo and CS datasets.
We postulate this is because the node attributes usually have high dimensions, which may lead to the model's overfitting.
In the following~\autoref{section:analysis}, we demonstrate that besides enabling high AUC attack performance, the posteriors also provide high robustness for link stealing attacks.
We attribute these two favorable properties to the aggregation function of the GNN, as it capitalizes fully on the neighbor node attributes and graph structures.
Due to space limitations, we present the attack performance on the other three target models in~\autoref{appendix:ablation_diff_target}, and the results demonstrate that our attacks are still effective.
We also explore different layers of attack models in \autoref{appendix:ablation_diff_attack_model}, showing that the attacks are in general effective.

\mypara{Different Shadow Models' Architectures}
We then investigate the variants of different architectures of shadow models.
We present the attack performance when the target model is GraphSAGE, and the shadow model's architecture is one of the four architectures mentioned in \autoref{section:preliminary}.
As shown in~\autoref{table:attack_performance_2}, we only present the results of the \post attacks on all six datasets since the different architectures only affect the posteriors.
We find those shadow models with different architectures still yield excellent attack performance.
We can also observe that the choice of the shadow model's architecture has a more significant impact on datasets with larger densities such as Photo and CS.
As mentioned in~\autoref{section:setup}, these datasets have more complex neighbor relationships, so different aggregating methods can lead to a larger variance in the attack performance of the \post attacks.
We calculate the gap between the best and worst attack performance of each \post attack and take the average of all \post attacks' gaps on each dataset.
The gap is negligible on sparse datasets like DBLP and Pubmed.
For example, the average gap is 0.55\% on the DBLP and Pubmed datasets.
The gap becomes larger on dense datasets such as CS and Photo.
For instance, the average gap is 12.67\% on the Photo dataset.
Since node attributes and graph features do not depend on the shadow model's architecture, attackers can additionally rely on the \comb attacks when they have no knowledge about the target model's architecture.
We report the results of the \comb attacks in~\autoref{appendix:ablation_diff_shadow}.
As we can see, the \comb attacks are architecture-agnostic.
For example, the average gap is only 3.2\% AUC using the \comb attacks on the Photo dataset, and the gap is only 1.5\% in Attack-5.
Overall, shadow models with architecture different from the target model still achieve excellent attack performance.
Hence, we can conclude that it is not necessary for the adversary to have knowledge of the target model's architecture as the choice of the shadow model's architecture does not have a significant impact on the attack performance.

\begin{table}[!t]
\caption{Effect of the different distributions of the shadow datasets on the attack performance.
Both the target and shadow architectures are GraphSAGE.
The evaluation metric is the average AUC score of five runs.}
\label{table:attack_performance_dist}
\centering
\renewcommand{\arraystretch}{1.1}
\scalebox{0.85}{
\begin{tabular}{c|c c c c c c}
\toprule
Target & \multicolumn{6}{c}{Shadow Dataset}  \\
Dataset & Cora & Pubmed  & DBLP & Photo & CS & LastFM \\
\midrule
Cora & 0.854 & 0.847 & \textbf{0.867} & 0.856 & 0.862 & 0.849 \\
Pubmed & 0.737 & \textbf{0.814} & 0.756 & 0.768 & 0.763 & 0.752 \\
DBLP & 0.782 & 0.805 & \textbf{0.837} & 0.788 & 0.794 & 0.785 \\
Photo & 0.863 & \textbf{0.881} & 0.879 & 0.876 & 0.874 & 0.872 \\
CS & 0.845 & 0.834 & 0.834 & 0.840 & \textbf{0.841} & 0.842 \\
LastFM & 0.881 & 0.882 & 0.876 & 0.882 & 0.880 & \textbf{0.882} \\
\bottomrule
\end{tabular}
}
\end{table}

\mypara{Different Shadow Datasets' Distributions}
In previous experiments, we leverage the shadow datasets from the same distribution as the target dataset to train the shadow models.
Here, we relax the assumption by leveraging different distribution shadow datasets.
To avoid the dimension mismatch problem, we redesign input features over the variable-length posteriors.
Specifically, we calculate the entropy with pairwise functions for posteriors pairs and combine the result vectors with cosine similarity, JS divergence, and correlation distance.
These metrics measure the similarity between posteriors pairs.
As shown in~\autoref{table:attack_performance_dist}, the results on the diagonal are from datasets with the same distribution but with different input features, demonstrating that our newly designed input features are effective.
We focus on Attack-1, as it only uses limited knowledge about graph structures but achieves a high attack success rate.
We can observe that, with shadow datasets from dissimilar distributions, the \post attacks can still achieve similar results on most datasets, even slightly better.
When the target model is trained on sparse datasets like Pubmed and DBLP, the attack performance is more susceptible to the distribution of the shadow dataset.
Overall, it demonstrates that information leaked from the GNN models can be transferred across datasets with dissimilar distributions.

\mypara{Different Shadow Datasets' Sizes}
So far, we assume the size of the shadow dataset and the target dataset are the same.
However, it is likely that the attacker has no knowledge about the size of the target dataset.
Therefore, we investigate the influence of shadow datasets' sizes on attack performance.
Following the above experiment, we focus on Attack-1.
We sample nodes in different proportions $\{10\%, 20\%, 30\%, 50\%, 100\%\}$ from the original shadow dataset to construct new shadow datasets.
The results are summarized in \autoref{table:attack_performance_size}.
We can observe that the attack performance is almost the same when the dataset size changes from 20\% to 100\%.
Even with only 10\% dataset, the attack performance is still acceptable.
Note that the proportion is divided by the number of nodes, and the corresponding number of edges is much smaller than the proportion.
For example, the subgraph that contains 10\% nodes only has 2.88\% edges on the LastFM dataset, indicating our attacks are still effective with limited access to the same distribution shadow dataset.
The reason is that the shadow model can still learn the ``correct'' predictions with a smaller dataset; hence the adversary can obtain the accurate similarities of the posteriors between node pairs.
In conclusion, our attacks can be carried out effectively using a much smaller shadow dataset than the target dataset, which makes the attack more practical.

\mypara{Takeaways}
First, we show that when the adversary has no knowledge about the graph structures, our attacks can achieve outstanding performance, e.g., outperforming the baselines by 14.2\% AUC on DBLP.
It demonstrates that posteriors leak rich information, even with no knowledge about the graph structures, to steal links from the inductive GNN models.
Also, our attacks still work well even with limited neighborhood information (i.e., attacks with 1-hop query perform similarly as those with 2-hop query).
Second, when the adversary has both node attributes and graph structures, the \comb attacks can achieve higher AUC scores than baselines on all datasets, which indicates the inductive GNN models leak extra information to enhance the link stealing attack.
Our attacks outperform the baselines by an average of 6.5\% AUC (on DBLP, it is 12.5\% higher).
Third, the experiments show that we can relax the assumptions about the shadow models' architectures, shadow datasets' distributions, and shadow datasets' sizes, as they do not have a significant impact on the attack performance.
This advantageous property makes our attacks a serious threat against inductive GNN models in real-world scenarios.

\begin{table}[!t]
\caption{Effect of the shadow datasets' sizes.
Both the target and shadow architectures are GraphSAGE.
The target and shadow datasets are the same.
The average AUC score of five runs is reported.}
\label{table:attack_performance_size}
\centering
\renewcommand{\arraystretch}{1.1}
\scalebox{0.85}{
\begin{tabular}{c|c c c c c}
\toprule
& \multicolumn{5}{c}{Dataset Size}  \\
Dataset &   100\% &    50\% &    30\% &    20\% &    10\% \\
\midrule
Cora        & \textbf{0.849} & 0.833 & 0.839 & 0.827 & 0.789 \\
Pubmed      & \textbf{0.798} & 0.776 & 0.770 & 0.762 & 0.742 \\
DBLP        & \textbf{0.823} & 0.819 & 0.816 & 0.818 & 0.761 \\
Photo       & \textbf{0.902} & 0.895 & 0.901 & 0.890 & 0.869 \\
CS          & \textbf{0.840} & 0.839 & 0.825 & 0.832 & 0.823 \\
LastFM      & 0.869 & 0.878 & 0.870 & \textbf{0.884} & 0.727 \\
\bottomrule
\end{tabular}
}
\end{table}

\begin{figure*}[!t]
\centering
\begin{subfigure}{1.8\columnwidth}
\includegraphics[width=0.7\columnwidth]{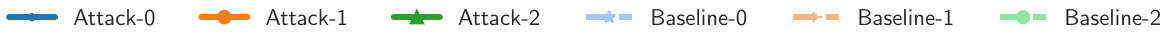}
\end{subfigure}
\begin{subfigure}{1.8\columnwidth}
\includegraphics[width=\columnwidth]{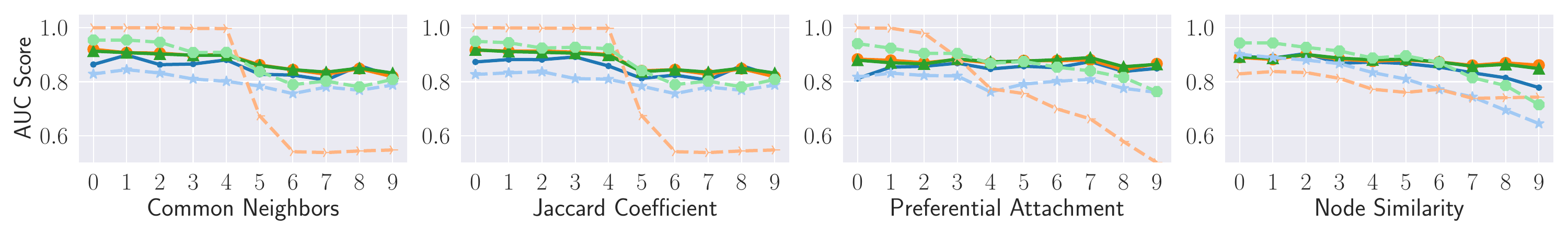}
\end{subfigure}
\begin{subfigure}{1.8\columnwidth}
\includegraphics[width=0.7\columnwidth]{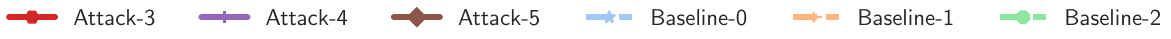}
\end{subfigure}
\begin{subfigure}{1.8\columnwidth}
\includegraphics[width=\columnwidth]{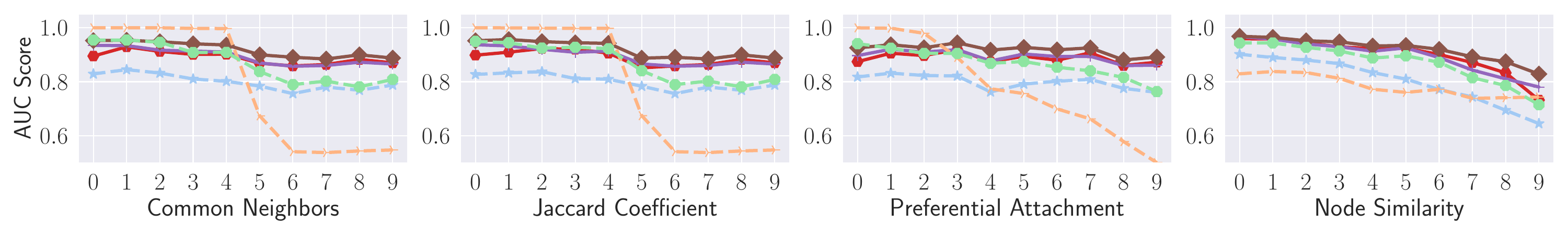}
\end{subfigure}
\begin{subfigure}{1.8\columnwidth}
\includegraphics[width=0.6\columnwidth]{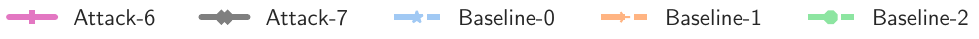}
\end{subfigure}
\begin{subfigure}{1.8\columnwidth}
\includegraphics[width=\columnwidth]{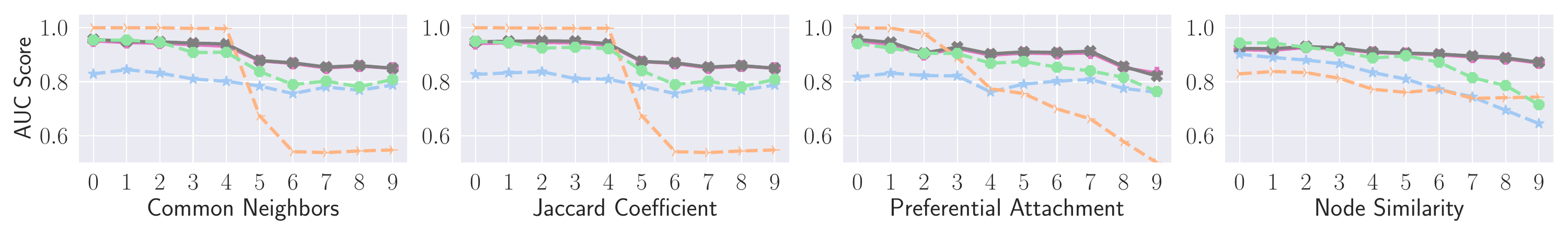}
\end{subfigure}
\begin{subfigure}{1.8\columnwidth}
\includegraphics[width=0.6\columnwidth]{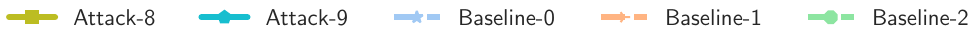}
\end{subfigure}
\begin{subfigure}{1.8\columnwidth}
\includegraphics[width=\columnwidth]{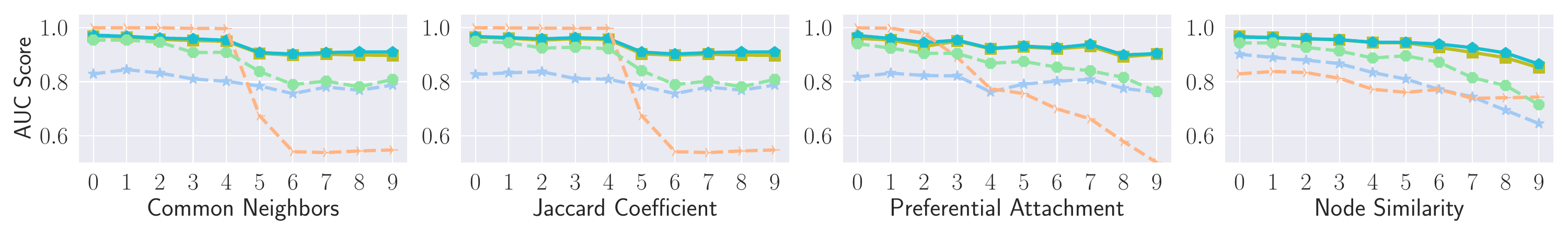}
\end{subfigure}
\caption{AUC score for our attacks and baselines on ten different groups on the LastFM dataset.
The ten groups are formed by categorizing all existing links in $G_{\textit{Target}}^{\textit{Train}}$ (the positive pairs in $D_{\textit{Attack}}^{\textit{Test}}$) based on node attributes and three graph metrics, respectively.
The x-axis represents different groups in descending order of their corresponding metric values.
The y-axis represents the AUC scores.
Each column represents one metric.
The first to fourth rows are $A_{(p, ~\cdot, ~\cdot)}^{*}$, $A_{(p, n, ~\cdot)}^{*}$, $A_{(p, ~\cdot, g)}^{*}$, and $A_{(p, n, g)}^{*}$, respectively.
We report results on the other five datasets in~\autoref{appendix:robust}, and similar conclusions can be drawn.}
\label{figure:robustness_lastfm}
\end{figure*}

\subsection{Fine-Grained Analysis}
\label{section:analysis}

\mypara{Robustness} 
Previous work~\cite{LK07} shows that graph features based on node neighborhoods contribute to good link prediction performance, which is in line with the intuition that a node pair is more likely to be linked if their neighbors have a large overlap.
Also, due to the intrinsic similarity of neighbor node attributes, traditional link prediction methods that use only node attributes can also be successful~\cite{GTMHSSSS14}.

Such graph features and node attributes are also involved in our combined link stealing attacks.
Thus, we select three graph metrics and a node metric and investigate the correlation between these metrics and our attack performance.
Specifically, we investigate the robustness of the proposed attacks and baselines with respect to three graph metrics (common neighbors, preferential attachment, and Jaccard coefficient) and one node metric (node similarity).
We categorize all existing links in the $G_{\textit{Target}}^{\textit{Train}}$ (the positive pairs in $D_{\textit{Attack}}^{\textit{Test}}$) into ten groups based on these four metrics, respectively.
We then calculate the AUC score for each group by combining node pairs in the group with all negative pairs in $D_{\textit{Attack}}^{\textit{Test}}$, as the AUC score is insensitive to imbalanced classification.
As illustrated in~\autoref{figure:robustness_lastfm}, we compare our attacks and baselines in terms of AUC scores on different groups sorted by four metrics on the LastFM dataset.
We observe that the AUC score for Baseline-1 drops rapidly when the graph metric value decreases (group index increases), i.e., lower graph features result in worse attack performance.
We also find that on decreasing the values of node similarity, Baseline-0/2 has a significant deterioration in the attack performance, especially in the latter groups.
It indicates that it is hard for the baseline attacks to distinguish between positive and negative pairs with either lower graph features' values or low node similarity, as the attack model tends to classify all of them as negative pairs.
Although Baseline-0 and Baseline-2 do not show a substantial decreasing trend in the groups sorted by graph metrics, their AUC scores are relatively low in all ten groups.
A similar situation exists for Baseline-1: as node similarity decreases, Baseline-1 is not significantly affected, but its overall attack performance is relatively low.

As for \post attacks $A_{(p, ~\cdot, ~\cdot)}^{*}$, they have a stable performance on all groups sorted by either graph metrics or node similarity, showing the posteriors itself can enable robust link stealing attacks.
As for \comb attacks $A_{(p, n, ~\cdot)}^{*}$, $A_{(p, ~\cdot, g)}^{*}$, and $A_{(p, n, g)}^{*}$, they have high robustness with respect to different graph features and different node similarities and outperform baseline attacks by a large margin, showing that posteriors can help distinguish positive pairs with lower metric values.
In other words, combining the posteriors, inherent graph features, and node attributes can result in more robust and better attack performance, especially in groups with lower metric values.
Overall, these observations demonstrate that information leaks from inductive GNNs, i.e., posteriors, can improve both the attack performance and the robustness of link stealing attacks with respect to different graph features and different node similarities.
We also show the comparison between our attacks and baselines on the other five datasets in~\autoref{appendix:robust}, and similar conclusions as above can be drawn.
In addition, the analysis of robustness
using Pearson correlation coefficient is also in~\autoref{appendix:robust}.

\begin{figure}[!t]
\centering
\begin{subfigure}{\columnwidth}
\includegraphics[width=\columnwidth]{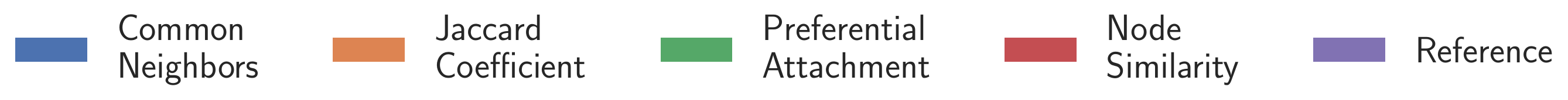}
\end{subfigure}
\begin{subfigure}{0.48\columnwidth}
\includegraphics[width=\columnwidth]{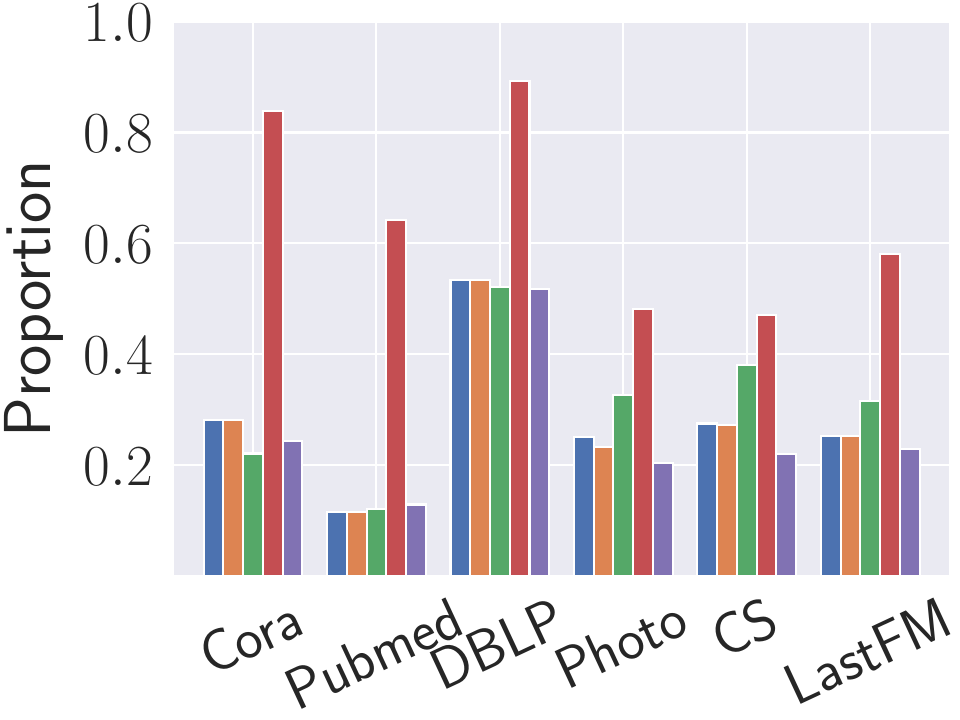}
\end{subfigure}
\begin{subfigure}{0.48\columnwidth}
\includegraphics[width=\columnwidth]{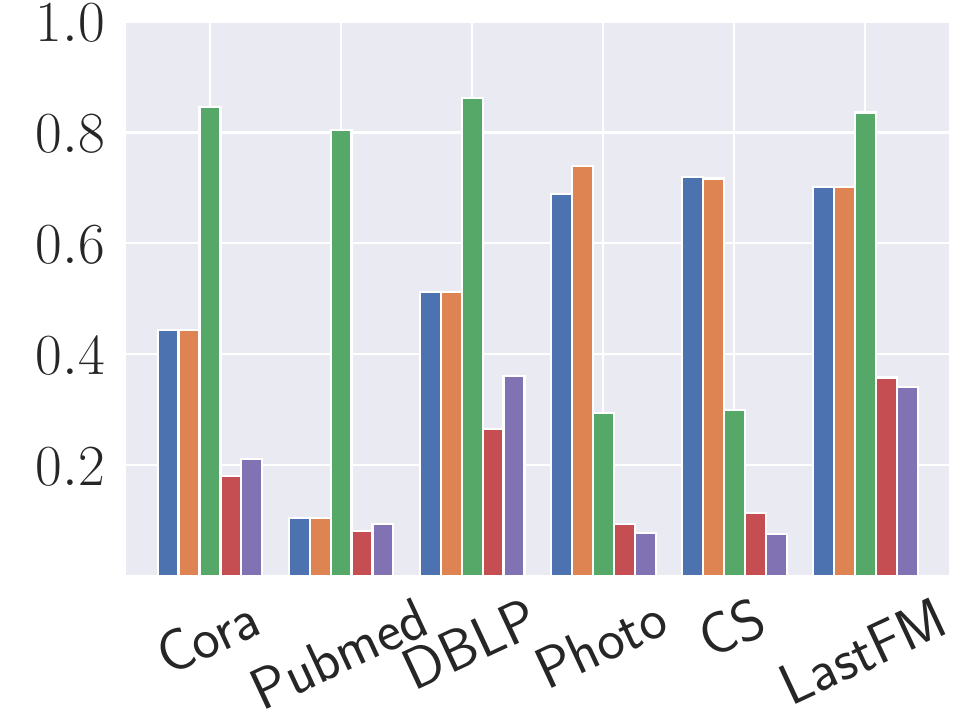}
\end{subfigure}
\caption{The proportion of the ``surprising'' links in the last group that is detected by Attack-1 but not by Baseline-0 (left figure) and by Attack-1 but not by Baseline-1 (right figure).}
\label{figure:s_link}
\end{figure}

\mypara{``Surprising'' Links}
In previous experiments, we demonstrate that our attacks achieve high robustness and accuracy.
This conclusion drives us to explore if our attacks could detect ``surprising'' links, i.e., those links that are correctly predicted by our attacks but not by the baseline attacks.
Specifically, we focus on Attack-1, the most efficient \post attack, as we are interested in the difference between posteriors, node attributes, and graph features.
We again categorize all positive pairs evenly into ten groups of positive pairs by four metrics, respectively.
We only focus on the last groups where the metric values are the lowest.
In~\autoref{figure:s_link}, each bar group presents the proportion of ``surprising'' links in the last group on different datasets.
Different colors denote the last group with respect to different metrics.
We also use the proportion of ``surprising'' links in all positive pairs as a reference (purple).
We observe that in the left figure, the proportions in the last group on all datasets with respect to node similarity are higher than the reference, indicating that links with lower node similarity can be better detected by Attack-1 but not by Baseline-0.
We postulate this is because such ``surprising'' links are used by the aggregation function in the GNN training procedure, which may strengthen the memorization of such links.
We have a similar observation in the right figure, i.e., \post attacks are more robust than Baseline-1 against three different graph metrics.

\mypara{Takeaways}
In a nutshell, we find that our attacks are more robust with respect to different graph features and different node similarities than the \trad methods.
This favorable property of our attacks can help the adversary detect ``surprising'' links that cannot be detected by the baseline attacks.

\begin{figure*}[!t]
\centering
\begin{subfigure}{1.8\columnwidth}
\includegraphics[width=0.3\columnwidth]{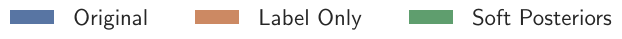}
\end{subfigure}
\begin{subfigure}{1.8\columnwidth}
\includegraphics[width=\columnwidth]{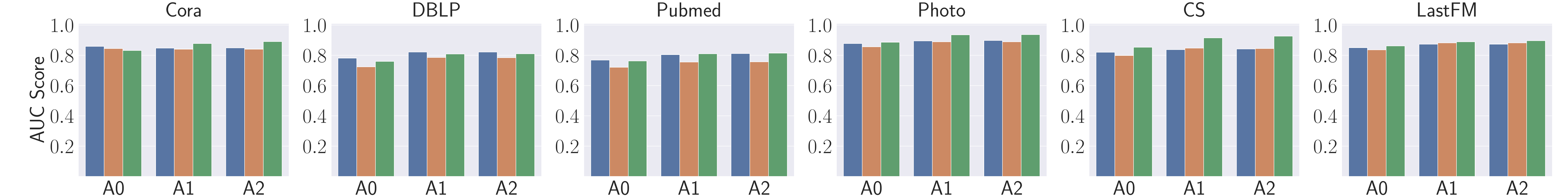}
\end{subfigure}
\caption{Attack performance of \post attacks under three scenarios.
We define different scenarios depending on the different outputs of the target model: original outputs generated by the softmax function (original), argmax outputs that only return the classification labels (label only), and soft outputs generated by softmax function with temperature scaling (soft posteriors).
Each bar group corresponds to a \post attack.
Each bar within a group represents a scenario.
The evaluation metric is the AUC score.}
\label{figure:label_only}
\end{figure*}

\subsection{Possible Defenses}

We investigate two defense mechanisms to mitigate link stealing attacks against inductive GNNs.

\mypara{Label-Only Outputs}
In this mechanism, the adversary can still perform queries with 0-hop, 1-hop, or 2-hop.
However, the target model only returns prediction labels rather than posteriors.
We assume the adversary has knowledge about the number of classes of the target model.
The adversary can also perform multiple queries to get enough labels to know the number of classes.
To avoid the node order issue, we convert two labels of a given node pair into two one-hot vectors and add them together as the attack input feature.
As we can see in~\autoref{figure:label_only}, our attack performance has decreased after applying the label-only defense, especially on the DBLP and Pubmed datasets.
The decrease in attack performance on DBLP and Pubmed is at most $5.8\%$.
In addition, there is only a slight fluctuation on the other four datasets, which may be due to network homophily that states that nodes that have similar attributes/labels are more likely to be linked together.
As illustrated in~\autoref{appendix:cdf}, we also observe that the leading probability of the target model's outputs dominates the rest, especially in Attack-1 and Attack-2.
This indicates the posteriors reveal almost the same amount of information as the labels, thus explaining why the link stealing attacks are still effective.
We next explore whether our attacks still work when the leading probability is tight with the second-largest probability.
Specifically, we leverage the ``softmax temperature'' technique~\cite{HVD15} to make the posteriors softer.
The temperature parameter $T$ is set to 20.
As illustrated in~\autoref{figure:label_only}, the attack performance increases rather than decreases on most datasets.
We attribute this observation to the fact that the soft posteriors provide more information in different classes that can boost the link stealing attacks.
Overall, our attacks can maintain good performance while dealing with different types of outputs.

\begin{figure*}[!t]
\centering
\begin{subfigure}{1.8\columnwidth}
\includegraphics[width=0.3\columnwidth]{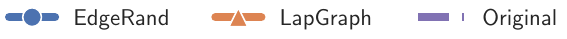}
\end{subfigure}
\begin{subfigure}{1.8\columnwidth}
\includegraphics[width=\columnwidth]{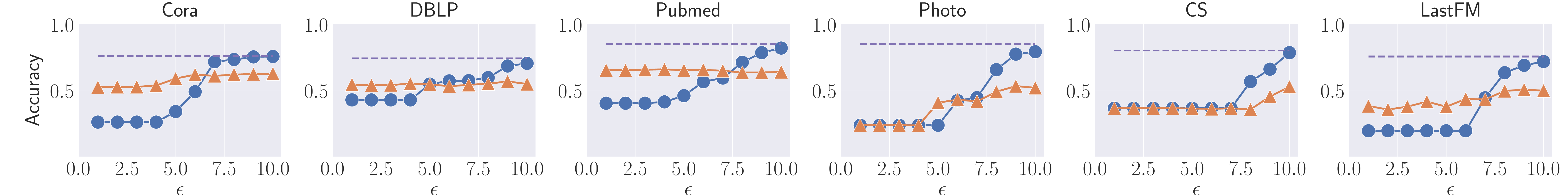}
\caption{Target Performance}
\label{figure:dp_target}
\end{subfigure}
\begin{subfigure}{1.8\columnwidth}
\includegraphics[width=\columnwidth]{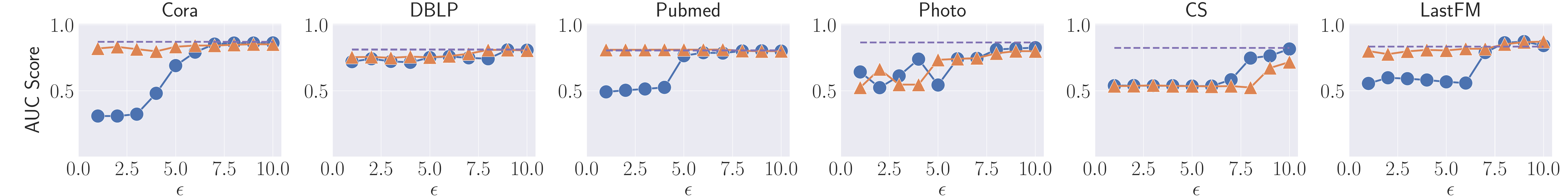}
\caption{Attack Performance}
\label{figure:dp_attack}
\end{subfigure}
\caption{(a) Target performance and (b) attack performance under varying privacy budgets $\epsilon \in [1, 10]$, when applying \edgerand and \lapgraph as defense mechanisms.
Each column corresponds to a dataset.
The x-axis denotes the privacy budget.
The y-axis represents (a) target performance measured by accuracy and (b) attack performance of Attack-1 measured by AUC score.
The dashed lines represent the (a) original target performance and (b) attack performance against vanilla GCN, while the solid lines represent the performance against two types of DP-GCN mechanisms.}
\label{figure:dp_defense}
\end{figure*}

\mypara{DP-GCN}
Another way to preserve the link privacy is to perturb the graph before training the GNN models.
Concretely, we leverage two differentially private graph convolutional network (DP-GCN) mechanisms~\cite{WLZL22}, i.e., \edgerand and \lapgraph.
DP-GCN guarantees inference results should be indistinguishable on any pair of neighboring input graphs differing in one edge by perturbing the adjacency matrix.
Specifically, the \edgerand mechanism randomly flips cells in the adjacency matrix based on a Bernoulli random variable.
The \lapgraph mechanism first computes the number of edges $T$ that needs to be kept in the perturbed graph using a small portion of the privacy budget and Laplace mechanism.
Then, it adds the Laplace noise to the entire adjacency matrix using the remaining privacy budget and keeps the largest $T$ entries in the perturbed graph.

We present target performance and attack performance on six datasets under privacy budget $\epsilon \in [1, 10]$, $\epsilon \in Z$ for two DP mechanisms, i.e., \edgerand and \lapgraph.
Following previous work~\cite{WLZL22}, both the target model and shadow model are GCN.
We focus on Attack-1, as it only uses limited knowledge about graph structures but achieves a high success rate.
The target performance, i.e., model utility, is reported in~\autoref{figure:dp_target} using the test accuracy as the metric, and the attack performance is reported in~\autoref{figure:dp_attack} using the test AUC score as the metric.

For \edgerand, we can observe that the attack performance and model utility decrease on all datasets via curtailing the privacy budgets since the privacy protection of the graph structure becomes stronger.
When the attack performance is close to a random guess, the model also becomes no longer usable.
It means that the attack success rate must be reduced at the cost of reducing target performance.
It is expected that the perturbed graph may create some ``fake'' links to mitigate the attacks, but such links may also affect the target model's training, which leads to lower utility.
Worse yet, the attack performance on the DBLP dataset does not drop significantly even at the cost of target performance.

For \lapgraph, we can only see a sharp drop in the attack performance on two Amazon datasets, i.e., the Photo dataset and the CS dataset.
The attack performance decreases along with the deterioration of model utility.
The model utility decreases significantly compared with the vanilla GCN on the other four datasets, while the attack performance has barely dropped.
Also, the model utility and attack performance do not decrease much with tightening privacy budgets.
It may be possible that as we continue to make the privacy budget tighter, the attack performance decreases.
Nevertheless, the target models are no longer usable when $\epsilon=1$.

\mypara{Takeaways}
To mitigate these attacks, we investigate a label-only mechanism and two state-of-the-art DP-GCN mechanisms, i.e., \edgerand and \lapgraph.
The results show that these defense mechanisms either have a negligible effect on the attack performance or decrease the attack performance while suffering unacceptable model utility reduction, indicating these defenses are impotent against our attacks.
We leave it as our future work to investigate more effective defenses.

\section{Related Work}

\mypara{Privacy Attacks Against GNNs}
This topic has gained momentum over the last few years.
Wu et al.~\cite{WYPY21} focus on graph-level membership inference attacks against GNNs trained for graph classification tasks.
Olatunji et al.~\cite{ONK21} propose a membership inference attack against node classification GNNs where the adversary's goal is to infer whether a specific node is used to train the target GNN model.
He et al.~\cite{HWWBSZ21} provide a complete attack taxonomy of node-level membership inference attacks by proposing three attacks where the adversary can perform a 0-hop, 1-hop, or 2-hop query based on different background knowledge.
Duddu et al.~\cite{DBS20} propose a membership inference attack in a white-box setting, where the adversary can leverage the intermediate outputs of the graph convolution layer, i.e., graph embeddings.
Besides membership inference, Zhang et al.~\cite{ZCBSZ22} also investigate privacy leakage of graph embeddings by mounting three inference attacks, i.e., property inference, subgraph inference, and graph reconstruction attacks.

There are two studies most relevant to our work.
He et al.~\cite{HJBGZ21} propose the first link stealing attack that aims to infer whether there exists a link between a given node pair in the target model's training graph.
Their work concentrates on the transductive setting, where the trained model has seen all testing data during the training phase, hence the users can only feed the identifiers of nodes to obtain prediction results.
It means that the posteriors derived from transductive GNNs utilize full neighbor information including the link the adversary intends to infer, regardless of whether the adversary possesses this neighbor information at inference time.
Our work shares similarities with~\cite{HJBGZ21} in using posteriors to design attack features.
However, we focus on the inductive setting where the trained model has not seen the testing data during the training phase (see details in~\autoref{section:gnn_setting}), and thus the adversary can only rely on limited and incomplete neighbor information, as the information of the link they intend to infer is missing, to construct nodes' subgraphs and then feed them into the target models to obtain posteriors.
The absence of complete neighbor information makes the link stealing attacks much more challenging.
Wu et al.~\cite{WLZL22} propose a query-based attack strategy, LinkTeller, which can recover edges among nodes of interest.
It focuses on a specific scenario where different data holders can host edge information and node attributes separately.
LinkTeller has two strong assumptions that are the key to the success of this work: (1) There is a model owner who is aware of the complete graph structure among nodes of interest and can obtain the complete neighbor information each time receiving a query; (2) The adversary is able to query the target model with the same set of nodes for multiple times.
In our work, we concentrate on the privacy issue of the target model's training graph by inferring whether there exists a link between two given nodes.
we also relax these assumptions in LinkTeller by (1) querying the target model with the adversary's own constructed subgraphs that have incomplete or even no neighbor information each time and (2) querying the target model once per node.
Our work stands in a more common scenario where the adversary needs to design attack input, such as nodes' subgraphs, based on their background knowledge.
There are some other works related to link stealing attacks, but they focus on different methodologies or properties.
Ding et al.~\cite{DDXF23} propose a graph poisoning attack to increase the effectiveness of link stealing attacks.
Zhang et al.~\cite{ZWWYXPY23} demonstrate that intra-class and inter-class node pairs have different levels of vulnerability to link stealing attacks.

\mypara{Other Attacks Against GNNs}
Recent studies have also shown that GNNs are vulnerable to other attacks such as model stealing attacks~\cite{DR19, WYPY20, SHHZ22}, backdoor attacks~\cite{ZJWG21, XPJW21} and adversarial attacks~\cite{BG192, CNKMPAV17, DLTHWZS18, WG19, WWTDLZ19, ZJWG21, ZAG18, ZG19, XCLCWHL19, EADP20, MDM20}.
Regarding the model stealing attacks, DeFazio et al.~\cite{DR19} propose a model extraction attack that can steal GNN models in the transductive setting.
Wu et al.~\cite{WYPY20} develop a series of model extraction attacks by categorizing the adversary's background knowledge along three dimensions, namely graph structures, node attributes, and shadow subgraphs.
He et al.~\cite{SHHZ22} propose model stealing attacks against inductive GNNs with six different attack scenarios.
The adversary's background knowledge is categorized into two dimensions, i.e., the target model's response and query graph.
Regarding the backdoor attacks, Zhang et al.~\cite{ZJWG21} propose the first backdoor attack against GNNs for graph classification tasks, which can be enabled by injecting predefined subgraphs.
Xi et al.~\cite{XPJW21} propose a backdoor attack against GNNs that the adversary can have no knowledge about the downstream tasks.
Regarding the adversarial attacks, Bojchevski et al.~\cite{BG192} focus on poisoning attacks on unsupervised node embeddings.
Chen et al.~\cite{CNKMPAV17} propose two adversarial attacks for graph-based clustering.
Dai et al.~\cite{DLTHWZS18} propose graph adversarial attacks in three settings, namely white-box, practical black-box, and restricted black-box settings.
Wang and Gong~\cite{WG19} propose adversarial attacks on collective classification methods via manipulating graph structures.
Wu et al.~\cite{WWTDLZ19} develop adversarial attacks by introducing the integrated gradient.
Zhang et al.~\cite{ZJWG21} propose the first subgraph-based backdoor attack on graph classification methods.
Z{\"u}gner et al.~\cite{ZAG18} propose an adversarial attack that can manipulate the node attributes and graph structures and preserve essential attributes of the graph at the same time.

\section{Conclusion}

In this paper, we comprehensively investigate the privacy leakage of the inductive graph neural networks through the lens of link stealing attacks.
Specifically, we propose two types of attacks, i.e., \post attack and \comb attack.
The threat model of the \post attack is defined by categorizing node topology, while that of the \comb attack is defined along three dimensions, i.e., posteriors, node attributes, and graph features.
We conduct extensive experiments on four popular GNN models over six real-world datasets.
The evaluation shows that inductive GNN leaks extra information to enable well-performing link stealing attacks even with weak assumptions.
Moreover, our investigations reveal that our attacks are robust with respect to different graph features and different node similarities.
To mitigate these attacks, we utilize label-only defense and the state-of-the-art DP-GCN mechanisms and show these defenses are impotent against our attacks.

\medskip
\mypara{Acknowledgements}
We thank all anonymous reviewers for their constructive comments.
This work is partially funded by the European Health and Digital Executive Agency (HADEA) within the project ``Understanding the individual host response against Hepatitis D Virus to develop a personalized approach for the management of hepatitis D'' (D-Solve) (grant agreement number 101057917).

\begin{small}
\bibliographystyle{IEEEtranS}
\bibliography{normal_generated_py3}    
\end{small}

\appendix

\section{GNN Architectures}
\label{appendix:gnn_models}

In this paper, we focus on four different GNN architectures, i.e., Graph Convolutional Network (GCN), GraphSAGE, Graph Attention Network (GAT), and Graph Isomorphism Network (GIN).

\mypara{GCN} 
Kipf et al.~\cite{KW17} propose GCN, which is the most famous and representative GNN method.
Combining message computation, aggregation, and update function, the layer of the GCN model can be defined as:
\begin{equation}
h_{v}^{(l)} = \sigma (\sum_{u \in N(v)} W^{(l)} \frac{h_{u}^{(l-1)}}{|N(v)|}),
\end{equation}
where the $W^{(l)}$ is the weight matrix and $\sigma$ represents the activation function.
The GCN layer sums up normalized messages of all neighbors and then applies the activation function to get the representations of node $\TargetNode$ in the layer $l$.
Note that the self-edges of node $\TargetNode$ are included in the summation.

\mypara{GraphSAGE} 
Hamilton et al.~\cite{HYL17} propose GraphSAGE, which can create inductive node embeddings for evolving graphs.
Several $\agg(\cdot)$ functions have also been proposed.
Furthermore, we leverage mean aggregation to take a weighted average of neighbors in this paper.
We formulate the GraphSAGE layer as follows:
\begin{equation}
h^{(l)}_{v} = \sigma (W^{(l)} \cdot \textit{CONCAT}(h^{(l-1)}_{v}, \sum_{u \in N(v)}  \frac{h^{(l-1)}_{u}}{|N(v)|})),
\end{equation}
where \textit{CONCAT} is the concatenate operation.

\mypara{GAT} 
Unlike GCN and GraphSAGE, where all neighbors are equally important to a node $v$, GAT computes attention weights $\alpha_{vu}$ to differentiate the information contribution between neighbors~\cite{VCCRLB18}.
We formulate the GAT layer as follows:
\begin{equation}
h^{(l)}_{v} = \sigma(\sum_{u \in N(v)}\alpha_{uv}W^{(l)}h_{u}^{(l-1)}),
\end{equation}
where $\alpha_{uv}$ and $W^{(l)}$ are learnable parameters.

\mypara{GIN} 
Xu et al.~\cite{XHLJ19} introduce GIN to provide a choice of $\agg(\cdot)$ and $\up(\cdot)$ that can make graph message passing neural networks equivalent to the Weisfeiler-Lehman (WL) algorithm.
The GIN layer is defined as:
\begin{equation}
h_{v}^{(l)} = \textit{MLP}^{(l)}((1 + \epsilon^{(l)}) \cdot h_{v}^{(l-1)} + \sum_{u \in N(v)} h^{(l-1)}_{u}),
\end{equation}
where $\epsilon^{(l)}$ is a learnable parameter and the multi-layer perceptions ($\textit{MLP}$) can represent the composition of functions.

\section{Grid Search}
\label{appendix:grid_search}

Below, we describe the search space of the hyper-parameters of the target, shadow, baseline, and attack models.

\noindent \emph{Target and shadow models.}
\begin{itemize}
\item learning rate: $\{0.1, 0.01, 0.001, 0.0001\}$
\item number of hidden neurons: $\{64, 128, 256\}$
\item dropout rate: $\{0.3, 0.5, 0.7\}$ 
\item optimizer: $\{Adam, SGD\}$
\item pair-wise operation: $\{ \textit{Hardmard}, \textit{Average}, \textit{Weighted-L1}, \\ \textit{Weighted-L2}, \textit{ALL}\}$
\end{itemize}

\noindent \emph{Baseline and attack models.}
\begin{itemize}
\item learning rate: $\{0.01, 0.001, 0.002, 0.003, 0.004, 0.005, 0.006, 0.007\}$
\item number of hidden layers: $\{2, 3, 4, 5\}$
\item number of hidden neurons: $\{4, 8, 16, 32, 64, 128, 256\}$
\item batch size: $\{128, 256, 512, 1024\}$
\item dropout rate: $\{0.3, 0.5\}$
\item optimizer: $\{Adam, SGD\}$
\item pair-wise operation: $\{ \textit{Hardmard}, \textit{Average}, \textit{Weighted-L1}, \\ \textit{Weighted-L2}, \textit{ALL}\}$
\end{itemize}

$\textit{ALL}$ denotes the concatenations of all four operations.

\section{Attack Success on Different Target Architectures}
\label{appendix:ablation_diff_target}

We conduct evaluations of \post attacks (Attack-0/1/2) on different target architectures, as the choice of GNN target architectures only affects the posterior features in~\autoref{table:attack_performance_appendix}.
The results demonstrate that the proposed attacks remain effective on GIN, GAT, and GCN target models.

\begin{table*}[ht]
\caption{Attack performance of \post attacks on all six datasets when the target model and the shadow model are the same.
The average AUC score of five runs is reported.}
\label{table:attack_performance_appendix}
\centering
\renewcommand{\arraystretch}{1.1}
\scalebox{0.85}{
\begin{tabular}{c|c|c c c c|c|c|c c c c}
\toprule
& & \multicolumn{4}{c|}{$M_{T}$ and $M_{S}$} & & & \multicolumn{4}{c}{$M_{T}$ and $M_{S}$}  \\
Dataset & Method & GraphSAGE & GIN & GAT & GCN & Dataset & Method & GraphSAGE & GIN & GAT & GCN \\
\midrule
\multirow{3}{*}{Cora} & Attack-0 &  0.859 & 0.866 & 0.882 & 0.893 & \multirow{3}{*}{Pubmed} & Attack-0 &  0.768 & 0.752 & 0.764 & 0.764  \\ 
& Attack-1 & 0.849 & 0.852 & 0.884 & 0.871 & & Attack-1 & 0.806 & 0.800 & 0.796 & 0.806 \\
& Attack-2 & 0.849 & 0.858 & 0.878 & 0.868 & & Attack-2 & 0.809 & 0.805 & 0.800 & 0.809 \\
\midrule
\multirow{3}{*}{DBLP} & Attack-0 & 0.781 & 0.748 & 0.783 & 0.761 & \multirow{3}{*}{Photo} & Attack-0 &  0.877 & 0.781 & 0.838 & 0.840  \\ 
& Attack-1 & 0.821 & 0.811 & 0.803 & 0.815 & & Attack-1 &  0.898 & 0.850 & 0.901 & 0.872 \\
& Attack-2 & 0.867 & 0.881 & 0.874 & 0.837 & & Attack-2 &  0.898 & 0.825 & 0.907 & 0.877\\
\midrule
\multirow{3}{*}{CS} & Attack-0 & 0.817 & 0.795 & 0.791 & 0.777 & \multirow{3}{*}{LastFM} & Attack-0 & 0.850 & 0.829 & 0.852 & 0.810  \\ 
& Attack-1 &  0.838 & 0.864 & 0.851 & 0.826 & & Attack-1 & 0.869 & 0.868 & 0.865 & 0.838 \\
& Attack-2 & 0.845 & 0.870 & 0.853 & 0.827  & & Attack-2 & 0.867 & 0.881 & 0.874 & 0.837 \\
\bottomrule
\end{tabular}
}
\end{table*}

\section{Impact of Different Attack Models}
\label{appendix:ablation_diff_attack_model}

We explore the impact of different attack models in~\autoref{table:attack_performance_layer_effect}.
Specifically, we consider different layers of attack models (MLP classifiers) for \post attacks.
We observe that the effectiveness of posterior-only attacks remains effective.

\section{Impact of Different Shadow Architectures}
\label{appendix:ablation_diff_shadow}

We report the performance of \comb attacks with different shadow model architectures in~\autoref{table:attack_performance_appendix_2}, and the results demonstrate that the mismatched shadow model architectures can still enable successful \comb attacks.

\begin{table*}[ht]
\caption{Attack performance of \post attacks on six datasets when the target model and the shadow model are GraphSAGE.
Different model layers are used in the attack model.
The AUC score is reported.}
\label{table:attack_performance_layer_effect}
\centering
\renewcommand{\arraystretch}{1.1}
\scalebox{0.85}{
\begin{tabular}{c|c|c c c c|c|c|c c c c}
\toprule
& & \multicolumn{4}{c|}{ \# Layers } & & & \multicolumn{4}{c}{ \# Layers }  \\
Dataset & Method & 2 & 3 & 4 & 5 & Dataset & Method & 2 & 3 & 4 & 5 \\
\midrule
\multirow{3}{*}{Cora} & Attack-0 & 0.858 &  0.859 &  0.863 &  0.861 & \multirow{3}{*}{Pubmed} & Attack-0 & 0.763 &  0.768 &  0.766 &  0.764 \\ 
& Attack-1 & 0.842 &  0.848 &  0.852 &  0.847 & & Attack-1 & 0.796 &  0.805 &  0.808 &  0.806 \\ 
& Attack-2 & 0.843 &  0.849 &  0.852 &  0.847& & Attack-2 & 0.803 &  0.812 &  0.814 &  0.813 \\
\midrule
\multirow{3}{*}{DBLP} & Attack-0 & 0.777 &  0.782 &  0.776 &  0.777  & \multirow{3}{*}{Photo} & Attack-0 & 0.880 &  0.877 &  0.874 &  0.868 \\ 
& Attack-1 & 0.818 &  0.821 &  0.822 &  0.822 & & Attack-1 & 0.899 &  0.891 &  0.878 &  0.887 \\ 
& Attack-2 & 0.819 &  0.822 &  0.824 &  0.824 & & Attack-2 & 0.898 &  0.895 &  0.885 &  0.870 \\
\midrule
\multirow{3}{*}{CS} & Attack-0 & 0.823 &  0.809 &  0.810 &  0.801 & \multirow{3}{*}{LastFM} & Attack-0 & 0.859 &  0.851 &  0.839 &  0.845 \\
& Attack-1 & 0.840 &  0.835 &  0.817 &  0.813 & & Attack-1 & 0.866 &  0.873 &  0.865 &  0.871 \\
& Attack-2 & 0.847 &  0.844 &  0.813 &  0.807 & & Attack-2 & 0.878 &  0.874 &  0.864 &  0.881 \\
\bottomrule
\end{tabular}
}
\end{table*}

\section{Robustness Analysis}
\label{appendix:robust}

\begin{figure*}[ht]
\centering
\begin{subfigure}{1.8\columnwidth}
\includegraphics[width=0.7\columnwidth]{figs/pos_pair_comp_posterior_legend.pdf}
\end{subfigure}
\begin{subfigure}{1.8\columnwidth}
\includegraphics[width=\columnwidth]{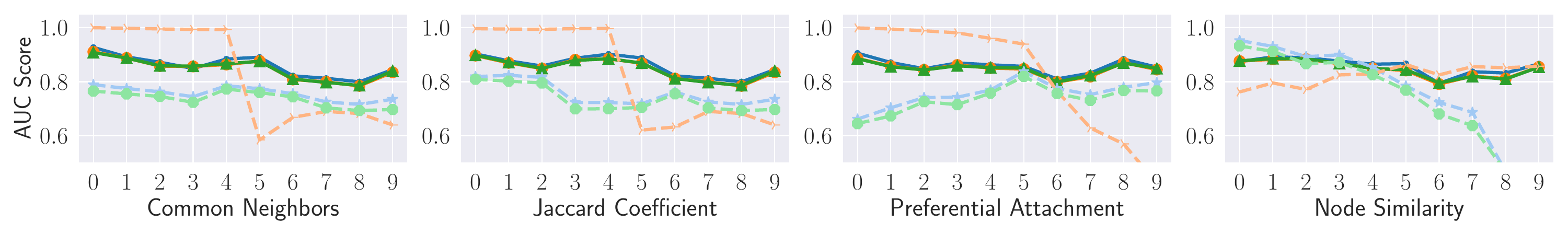}
\end{subfigure}
\begin{subfigure}{1.8\columnwidth}
\includegraphics[width=0.7\columnwidth]{figs/pos_pair_comp_node_legend.pdf}
\end{subfigure}
\begin{subfigure}{1.8\columnwidth}
\includegraphics[width=\columnwidth]{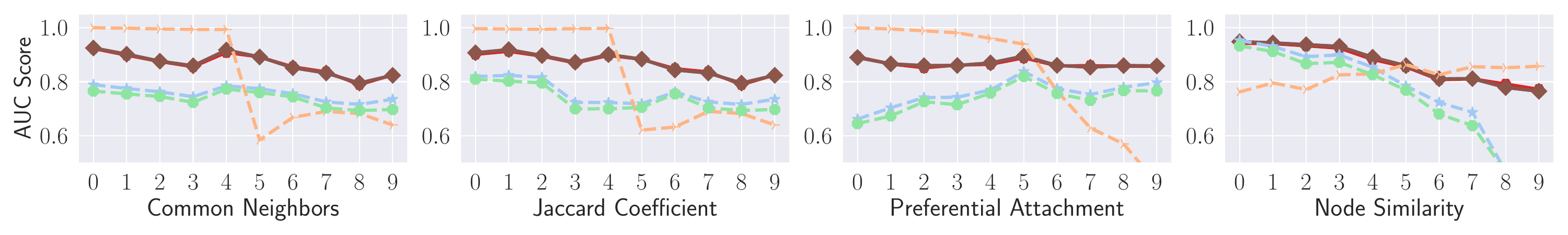}
\end{subfigure}
\begin{subfigure}{1.8\columnwidth}
\includegraphics[width=0.6\columnwidth]{figs/pos_pair_comp_graph_legend.pdf}
\end{subfigure}
\begin{subfigure}{1.8\columnwidth}
\includegraphics[width=\columnwidth]{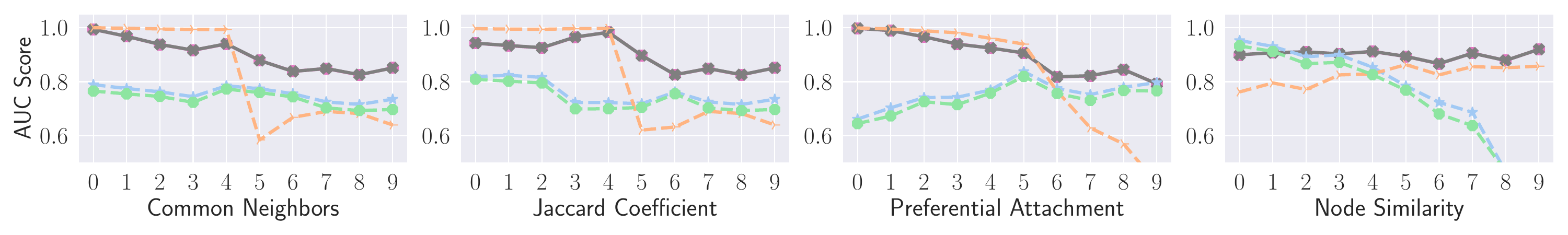}
\end{subfigure}
\begin{subfigure}{1.8\columnwidth}
\includegraphics[width=0.6\columnwidth]{figs/pos_pair_comp_node_graph_legend.pdf}
\end{subfigure}
\begin{subfigure}{1.8\columnwidth}
\includegraphics[width=\columnwidth]{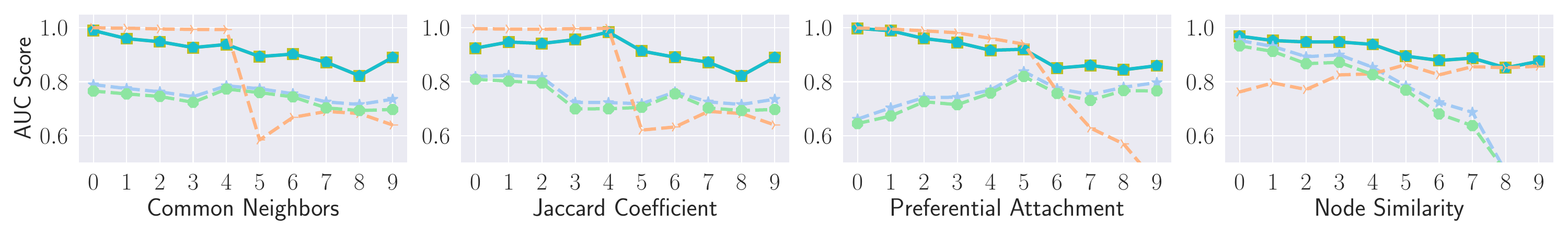}
\end{subfigure}
\caption{AUC score for our attacks and baselines on ten different groups on the Cora dataset.
The ten groups are formed by categorizing all existing links in $G_{\textit{Target}}^{\textit{Train}}$ (the positive pairs in $D_{\textit{Attack}}^{\textit{Test}}$) based on node attributes and three graph metrics, respectively.
The x-axis represents different groups in descending order of their corresponding metric values.
The y-axis represents the AUC scores.
Each column represents one metric.
The first to fourth rows are $A_{(p, ~\cdot, ~\cdot)}^{*}$, $A_{(p, n, ~\cdot)}^{*}$, $A_{(p, ~\cdot, g)}^{*}$, and $A_{(p, n, g)}^{*}$, respectively.}
\label{figure:robustness_cora}
\end{figure*}

\begin{figure*}[ht]
\centering
\begin{subfigure}{1.8\columnwidth}
\includegraphics[width=0.7\columnwidth]{figs/pos_pair_comp_posterior_legend.pdf}
\end{subfigure}
\begin{subfigure}{1.8\columnwidth}
\includegraphics[width=\columnwidth]{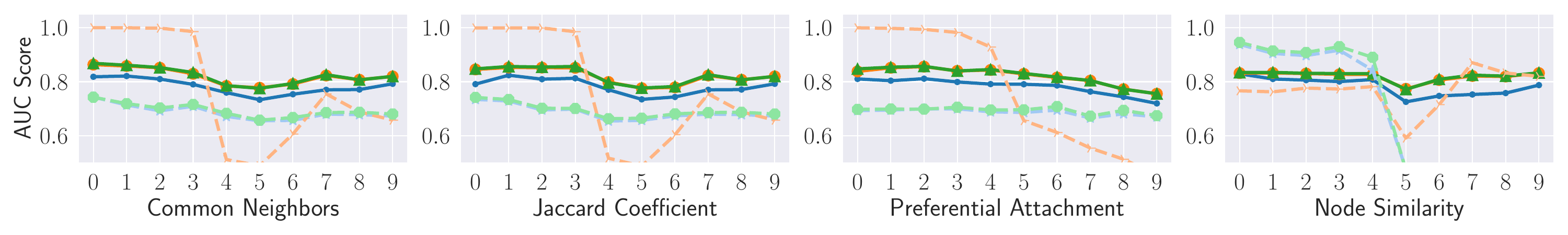}
\end{subfigure}
\begin{subfigure}{1.8\columnwidth}
\includegraphics[width=0.7\columnwidth]{figs/pos_pair_comp_node_legend.pdf}
\end{subfigure}
\begin{subfigure}{1.8\columnwidth}
\includegraphics[width=\columnwidth]{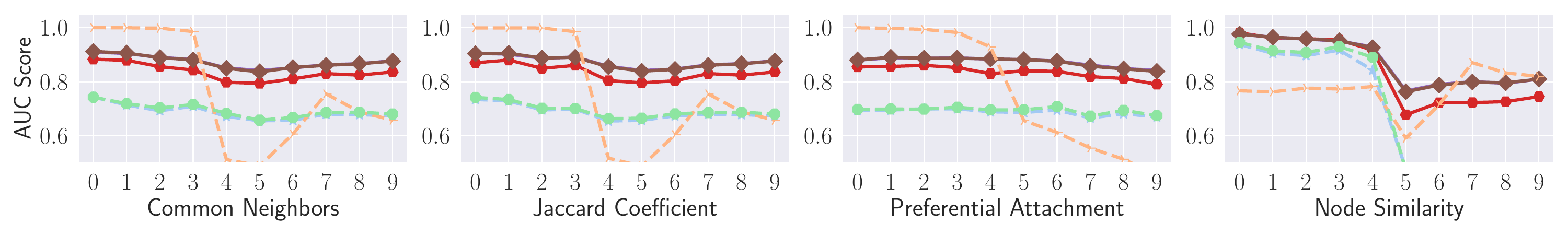}
\end{subfigure}
\begin{subfigure}{1.8\columnwidth}
\includegraphics[width=0.6\columnwidth]{figs/pos_pair_comp_graph_legend.pdf}
\end{subfigure}
\begin{subfigure}{1.8\columnwidth}
\includegraphics[width=\columnwidth]{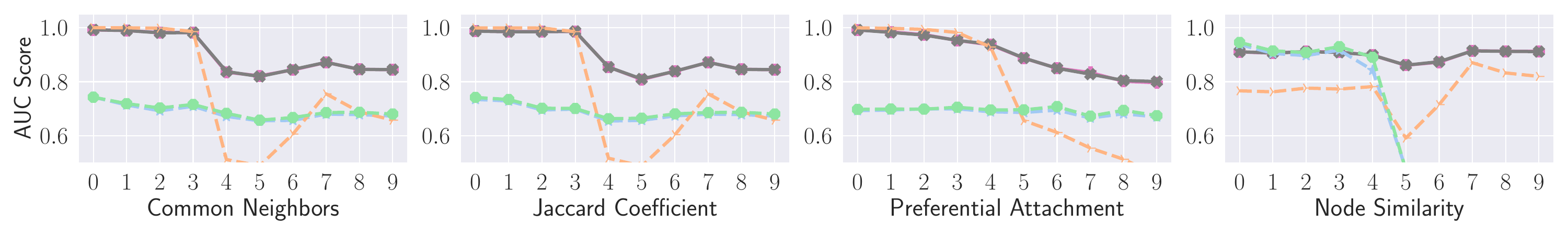}
\end{subfigure}
\begin{subfigure}{1.8\columnwidth}
\includegraphics[width=0.6\columnwidth]{figs/pos_pair_comp_node_graph_legend.pdf}
\end{subfigure}
\begin{subfigure}{1.8\columnwidth}
\includegraphics[width=\columnwidth]{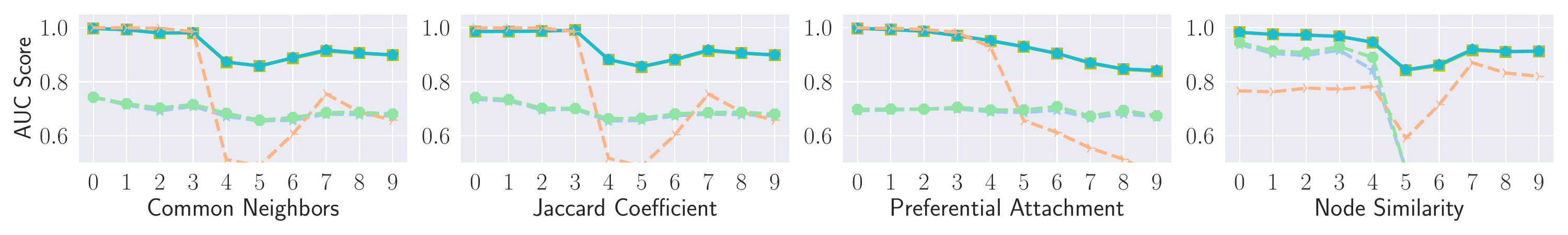}
\end{subfigure}
\caption{AUC score for our attacks and baselines on ten different groups on the DBLP dataset.
The ten groups are formed by categorizing all existing links in $G_{\textit{Target}}^{\textit{Train}}$ (the positive pairs in $D_{\textit{Attack}}^{\textit{Test}}$) based on node attributes and three graph metrics, respectively.
The x-axis represents different groups in descending order of their corresponding metric values.
The y-axis represents the AUC scores.
Each column represents one metric.
The first to fourth rows are $A_{(p, ~\cdot, ~\cdot)}^{*}$, $A_{(p, n, ~\cdot)}^{*}$, $A_{(p, ~\cdot, g)}^{*}$, and $A_{(p, n, g)}^{*}$, respectively.}
\label{figure:robustness_dblp}
\end{figure*}

\begin{figure*}[ht]
\centering
\begin{subfigure}{1.8\columnwidth}
\includegraphics[width=0.7\columnwidth]{figs/pos_pair_comp_posterior_legend.pdf}
\end{subfigure}
\begin{subfigure}{1.8\columnwidth}
\includegraphics[width=\columnwidth]{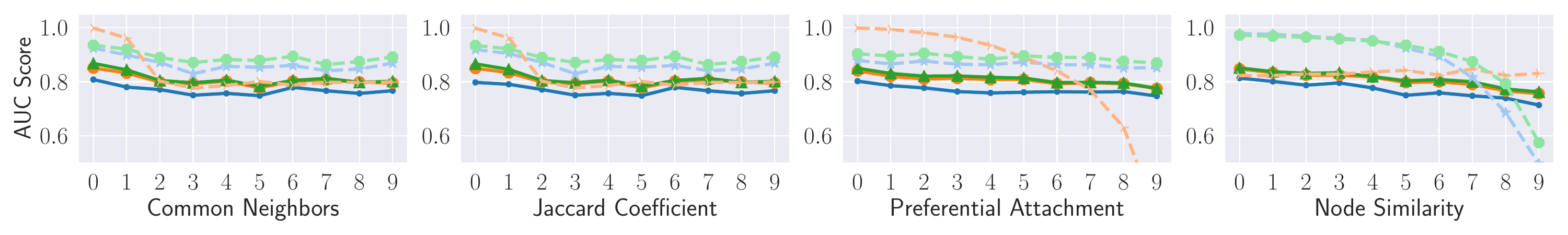}
\end{subfigure}
\begin{subfigure}{1.8\columnwidth}
\includegraphics[width=0.7\columnwidth]{figs/pos_pair_comp_node_legend.pdf}
\end{subfigure}
\begin{subfigure}{1.8\columnwidth}
\includegraphics[width=\columnwidth]{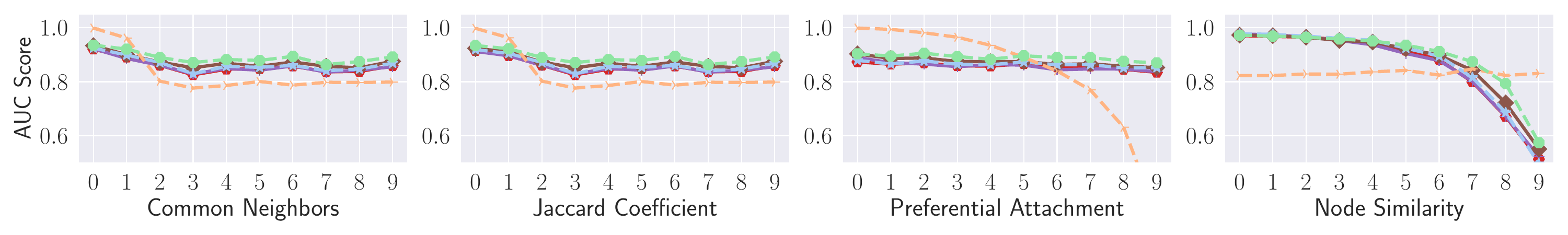}
\end{subfigure}
\begin{subfigure}{1.8\columnwidth}
\includegraphics[width=0.6\columnwidth]{figs/pos_pair_comp_graph_legend.pdf}
\end{subfigure}
\begin{subfigure}{1.8\columnwidth}
\includegraphics[width=\columnwidth]{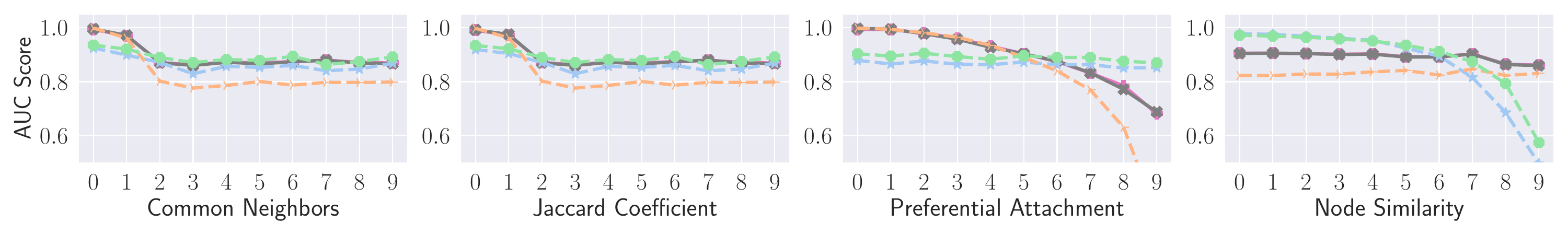}
\end{subfigure}
\begin{subfigure}{1.8\columnwidth}
\includegraphics[width=0.6\columnwidth]{figs/pos_pair_comp_node_graph_legend.pdf}
\end{subfigure}
\begin{subfigure}{1.8\columnwidth}
\includegraphics[width=\columnwidth]{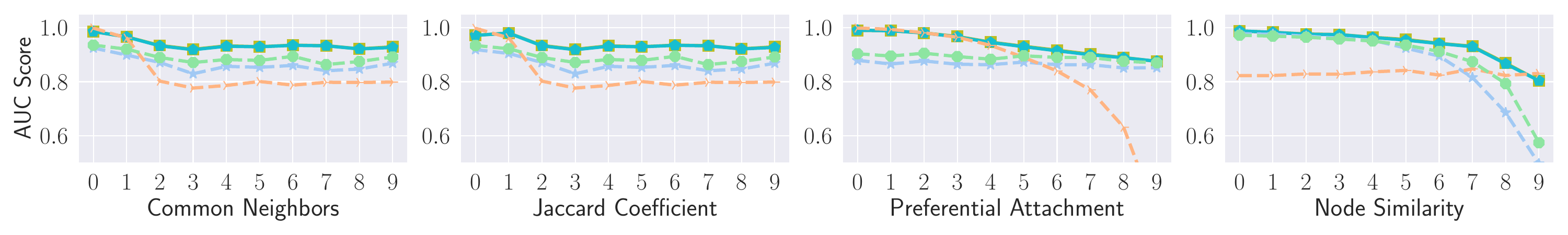}
\end{subfigure}
\caption{AUC score for our attacks and baselines on ten different groups on the Pubmed dataset.
The ten groups are formed by categorizing all existing links in $G_{\textit{Target}}^{\textit{Train}}$ (the positive pairs in $D_{\textit{Attack}}^{\textit{Test}}$) based on node attributes and three graph metrics, respectively.
The x-axis represents different groups in descending order of their corresponding metric values.
The y-axis represents the AUC scores.
Each column represents one metric.
The first to fourth rows are $A_{(p, ~\cdot, ~\cdot)}^{*}$, $A_{(p, n, ~\cdot)}^{*}$, $A_{(p, ~\cdot, g)}^{*}$, and $A_{(p, n, g)}^{*}$, respectively.}
\label{figure:robustness_pubmed}
\end{figure*}

\begin{figure*}[ht]
\centering
\begin{subfigure}{1.8\columnwidth}
\includegraphics[width=0.7\columnwidth]{figs/pos_pair_comp_posterior_legend.pdf}
\end{subfigure}
\begin{subfigure}{1.8\columnwidth}
\includegraphics[width=\columnwidth]{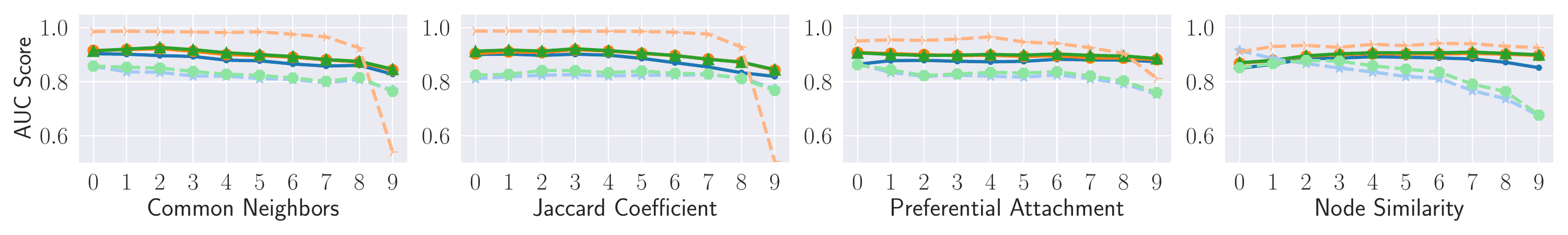}
\end{subfigure}
\begin{subfigure}{1.8\columnwidth}
\includegraphics[width=0.7\columnwidth]{figs/pos_pair_comp_node_legend.pdf}
\end{subfigure}
\begin{subfigure}{1.8\columnwidth}
\includegraphics[width=\columnwidth]{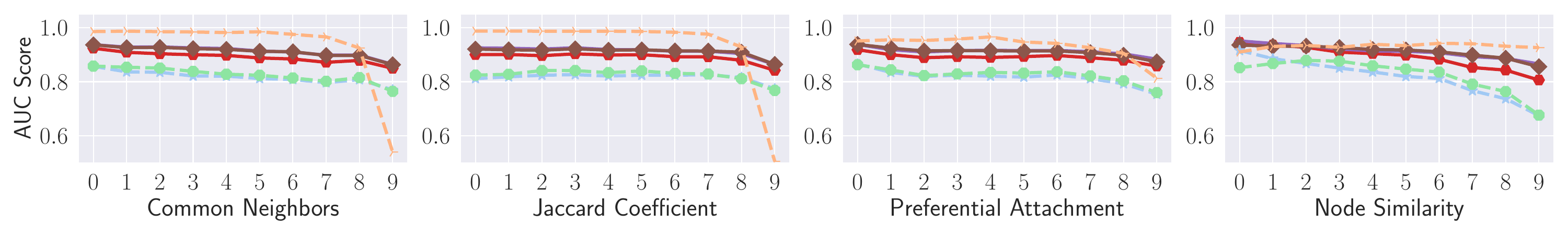}
\end{subfigure}
\begin{subfigure}{1.8\columnwidth}
\includegraphics[width=0.6\columnwidth]{figs/pos_pair_comp_graph_legend.pdf}
\end{subfigure}
\begin{subfigure}{1.8\columnwidth}
\includegraphics[width=\columnwidth]{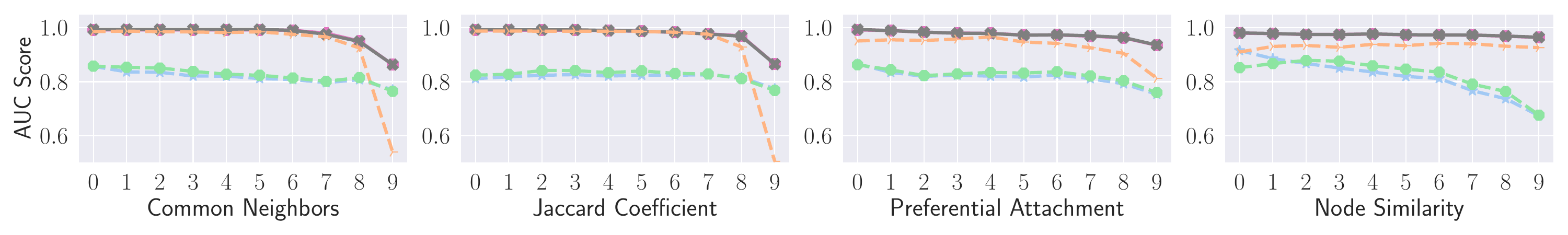}
\end{subfigure}
\begin{subfigure}{1.8\columnwidth}
\includegraphics[width=0.6\columnwidth]{figs/pos_pair_comp_node_graph_legend.pdf}
\end{subfigure}
\begin{subfigure}{1.8\columnwidth}
\includegraphics[width=\columnwidth]{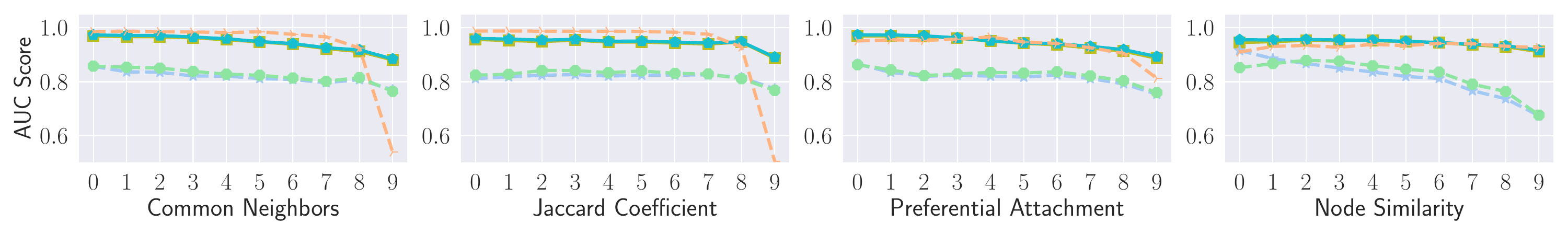}
\end{subfigure}
\caption{AUC score for our attacks and baselines on ten different groups on the Photo dataset.
The ten groups are formed by categorizing all existing links in $G_{\textit{Target}}^{\textit{Train}}$ (the positive pairs in $D_{\textit{Attack}}^{\textit{Test}}$) based on node attributes and three graph metrics, respectively.
The x-axis represents different groups in descending order of their corresponding metric values.
The y-axis represents the AUC scores.
Each column represents one metric.
The first to fourth rows are $A_{(p, ~\cdot, ~\cdot)}^{*}$, $A_{(p, n, ~\cdot)}^{*}$, $A_{(p, ~\cdot, g)}^{*}$, and $A_{(p, n, g)}^{*}$, respectively.}
\label{figure:robustness_photo}
\end{figure*}

\begin{figure*}[ht]
\centering
\begin{subfigure}{1.8\columnwidth}
\includegraphics[width=0.7\columnwidth]{figs/pos_pair_comp_posterior_legend.pdf}
\end{subfigure}
\begin{subfigure}{1.8\columnwidth}
\includegraphics[width=\columnwidth]{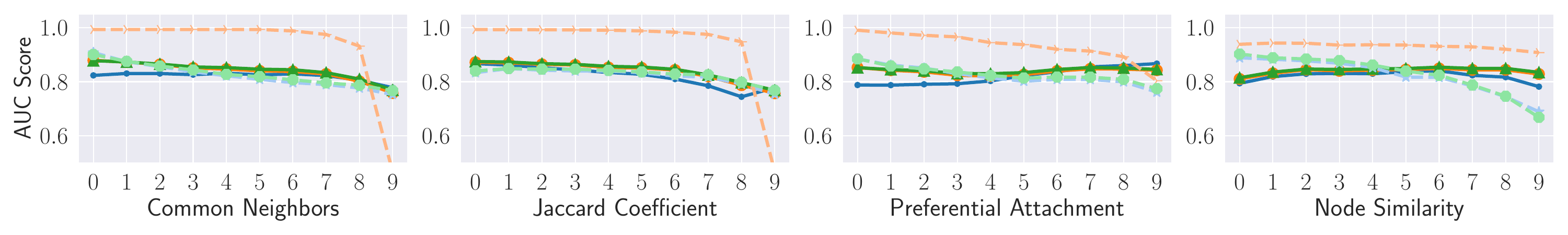}
\end{subfigure}
\begin{subfigure}{1.8\columnwidth}
\includegraphics[width=0.7\columnwidth]{figs/pos_pair_comp_node_legend.pdf}
\end{subfigure}
\begin{subfigure}{1.8\columnwidth}
\includegraphics[width=\columnwidth]{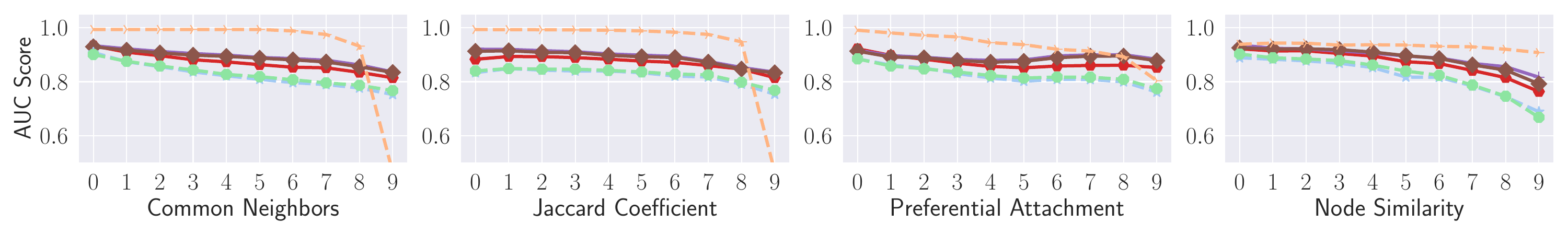}
\end{subfigure}
\begin{subfigure}{1.8\columnwidth}
\includegraphics[width=0.6\columnwidth]{figs/pos_pair_comp_graph_legend.pdf}
\end{subfigure}
\begin{subfigure}{1.8\columnwidth}
\includegraphics[width=\columnwidth]{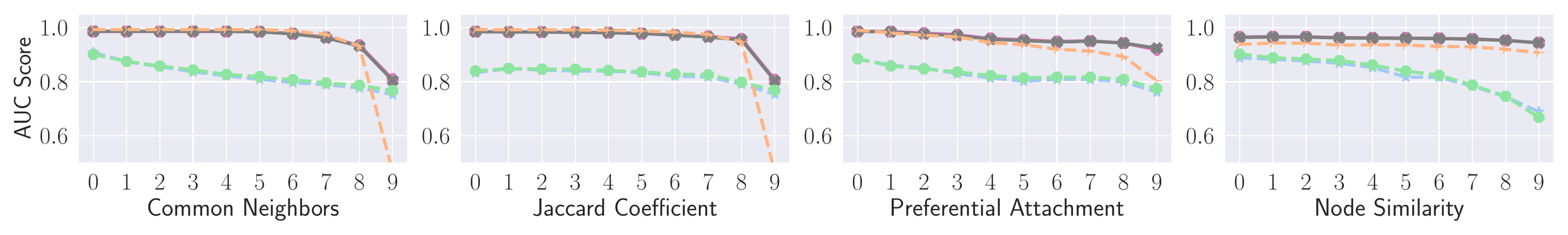}
\end{subfigure}
\begin{subfigure}{1.8\columnwidth}
\includegraphics[width=0.6\columnwidth]{figs/pos_pair_comp_node_graph_legend.pdf}
\end{subfigure}
\begin{subfigure}{1.8\columnwidth}
\includegraphics[width=\columnwidth]{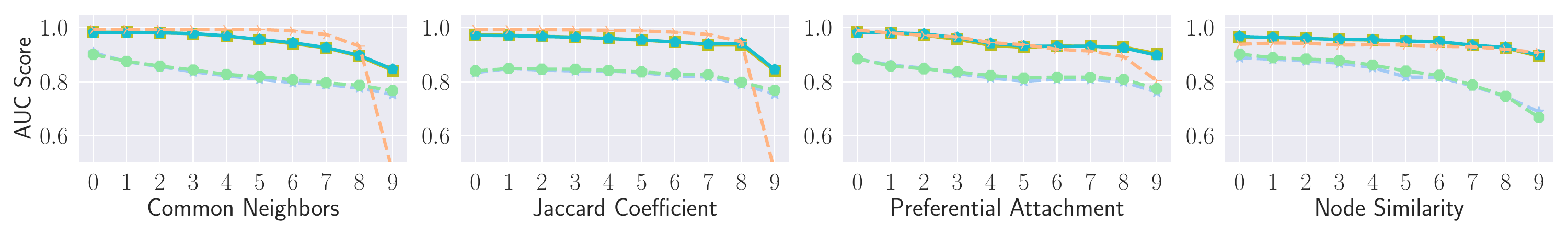}
\end{subfigure}
\caption{AUC score for our attacks and baselines on ten different groups on the CS dataset.
The ten groups are formed by categorizing all existing links in $G_{\textit{Target}}^{\textit{Train}}$ (the positive pairs in $D_{\textit{Attack}}^{\textit{Test}}$) based on node attributes and three graph metrics, respectively.
The x-axis represents different groups in descending order of their corresponding metric values.
The y-axis represents the AUC scores.
Each column represents one metric.
The first to fourth rows are $A_{(p, ~\cdot, ~\cdot)}^{*}$, $A_{(p, n, ~\cdot)}^{*}$, $A_{(p, ~\cdot, g)}^{*}$, and $A_{(p, n, g)}^{*}$, respectively.}
\label{figure:robustness_cs}
\end{figure*}

We compare our attacks and baselines in terms of AUC scores on different groups sorted by four metrics on the Cora (\autoref{figure:robustness_cora}), DBLP (\autoref{figure:robustness_dblp}), Pubmed (\autoref{figure:robustness_pubmed}), Photo (\autoref{figure:robustness_photo}), and CS (\autoref{figure:robustness_cs}) datasets.
We notice that the posteriors can enable robust link stealing attacks.
Also, combining the posteriors, inherent graph features and node attributes can result in more robust and better attack performance, especially in groups with lower metric values.

\section{Pearson Correlation Coefficient}
\label{appendix:pcc}

\begin{figure*}[ht]
\centering
\begin{subfigure}{1.8\columnwidth}
\includegraphics[width=\columnwidth]{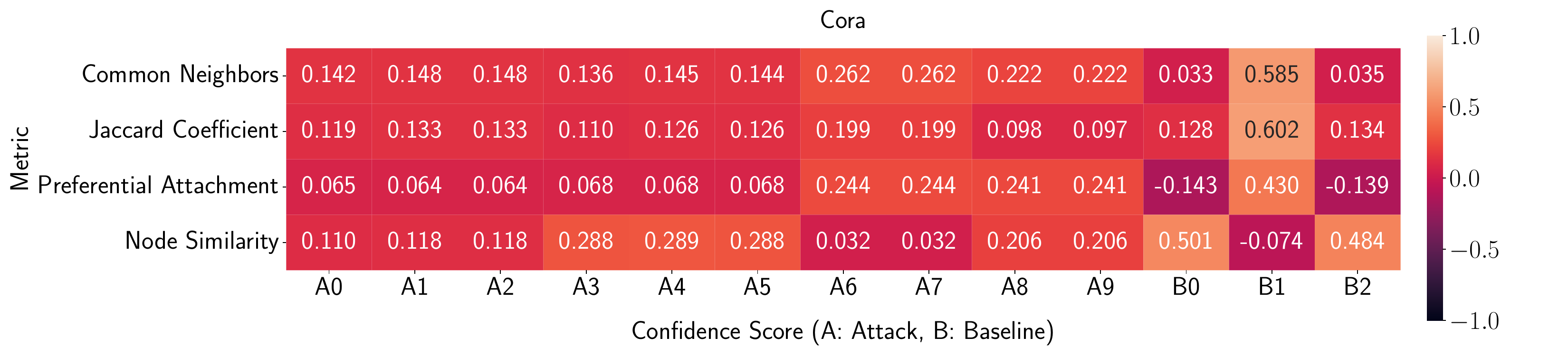}
\caption{Positive Pairs}
\end{subfigure}
\begin{subfigure}{1.8\columnwidth}
\includegraphics[width=\columnwidth]{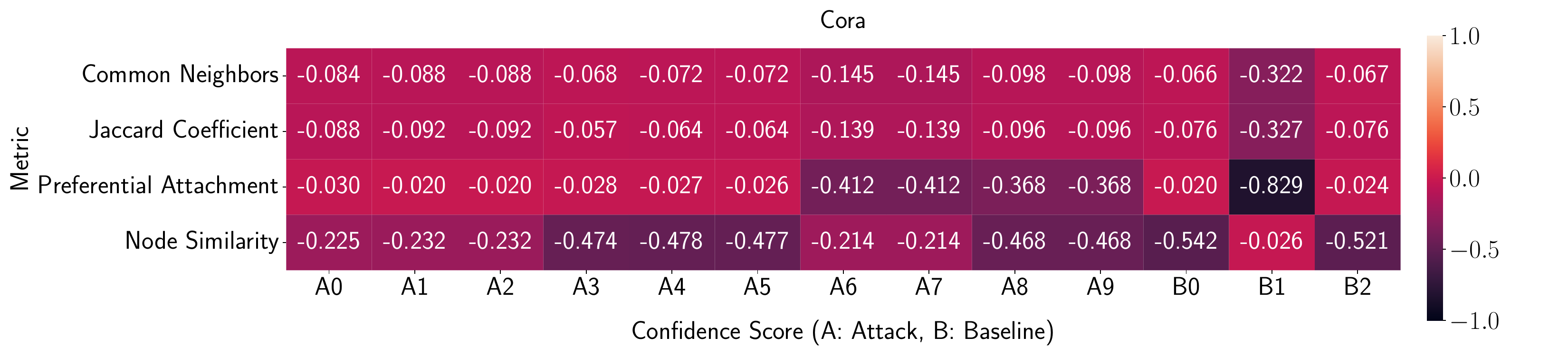}
\caption{Negative Pairs}
\end{subfigure}
\caption{Pearson correlation coefficient (PCC) between four metrics and attack performance.
We use PCC to measure the correlation between the attack performance and metric values.
The dataset is fixed to Cora.
The x-axis denotes ten link stealing attacks and three baselines.
The y-axis represents metrics of interest.
The value of PCC ranges from $-1$ to $1$.
Positive numbers represent positive correlations, while negative numbers indicate negative correlations.
The higher the absolute values of the PCC, the stronger are the correlations between the two variables.
We take the attack confidence scores to quantify the attack performance.}
\label{figure:heatmap_cora}
\end{figure*}

We also present the analysis of robustness using the Pearson correlation coefficient on the Cora dataset (\autoref{figure:heatmap_cora}).
The value of PCC ranges from $-1$ to $1$.
Positive numbers represent positive correlations, while negative numbers indicate negative correlations.
The higher the absolute values of the PCC, the stronger the correlations between two variables.
We take the attack confidence scores to quantify the attack performance.
As we can see, for positive pairs, there is a positive correlation between attack performance and these metrics in link stealing attacks, while there is a negative correlation for negative pairs.
The results are in line with our intuition that node pairs that have a large overlap in terms of neighbors and high node similarity are more prone to form links, and vice versa.
Baseline-0 and Baseline-2 which use node attributes have a high correlation between attack performance and node similarity.
Baseline-1 and Baseline-2 both use graph features to design attack input features.
The correlations between Baseline-1's attack performance and the three graph features are strong.
However, Baseline-2 has almost no correlation between the attack performance and the graph features.
We attribute this to the fact that the high-dimensional node attributes have more information to affect the attack performance.
Compared to baselines, the PCC values of all \post attacks are closer to 0, which indicates that the \post attacks have little correlation between attack performance, graph features, and node similarity.
As node attributes are used to design attack input features in Attack-3, Attack-4, Attack-5, Attack-8, and Attack-9, the correlation between node similarity and attack performance is increased.
However, our attacks' correlation with the node similarity metric is still weaker than the baselines that use node attributes.
Meanwhile, \comb attacks using graph features also yield a boost in the correlation between graph features and attack performance.
Similarly, our attacks' correlations with graph features are weaker than the baselines using graph features.
The strong correlation can indicate that the attack model overly depends on a specific feature to perform the classification task.
Thus, the attack model loses robustness against that feature.

\section{The Cumulative Distribution Function of Leading Probability}
\label{appendix:cdf}

\begin{figure*}[ht]
\centering
\begin{subfigure}{1.8\columnwidth}
\includegraphics[width=0.3\columnwidth]{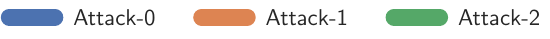}
\end{subfigure}
\begin{subfigure}{1.8\columnwidth}
\includegraphics[width=\columnwidth]{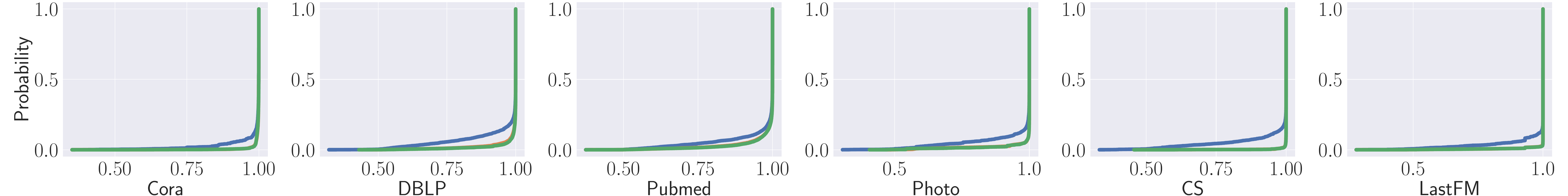}
\end{subfigure}
\caption{The cumulative distribution function of the leading probability of the target model's outputs.}
\label{figure:leading_posteriors}
\end{figure*}

The cumulative distribution function of leading probability is reported in~\autoref{figure:leading_posteriors}, we observe that the leading probability of the target model's outputs dominates the rest, especially in Attack-1 and Attack-2.

\begin{table*}[ht]
\caption{Attack performance of \comb attacks on all six datasets when the target model is GraphSAGE and the shadow model is one of four architectures we mentioned in \autoref{section:preliminary}.
The average AUC score of five runs is reported.}
\label{table:attack_performance_appendix_2}
\centering
\renewcommand{\arraystretch}{1.1}
\scalebox{0.85}{
\begin{tabular}{c|c|c c c c|c|c|c c c c}
\toprule
& & \multicolumn{4}{c|}{$M_{S}$} & & & \multicolumn{4}{c}{$M_{S}$}  \\
Dataset & Method & GraphSAGE & GIN & GAT & GCN & Dataset & Method & GraphSAGE & GIN & GAT & GCN \\
\midrule
\multirow{6}{*}{Cora} & Attack-3 &  0.876 & 0.874 & 0.880 & \textbf{0.881} & \multirow{6}{*}{Pubmed} & Attack-3 & 0.889 & \textbf{0.896} & 0.887 & 0.891 \\ 
& Attack-4 & 0.876 & 0.865 & \textbf{0.879} & \textbf{0.879} & & Attack-4 & 0.895 & 0.889 & \textbf{0.894} & 0.888 \\ 
& Attack-5 & 0.875 & 0.868 & 0.873 & \textbf{0.879} & & Attack-5 & \textbf{0.897} & 0.892 & 0.892 & 0.893 \\
& Attack-6 & 0.882 & 0.887 & 0.881 & \textbf{0.888} & & Attack-6 &  \textbf{0.881} & 0.878 & 0.877 & 0.876 \\
& Attack-7 & 0.884 & 0.880 & 0.881 & \textbf{0.888} & & Attack-7 & \textbf{0.892} &  0.891 &  0.890 &  0.889 \\
& Attack-8 & \textbf{0.882} & \textbf{0.882} & \textbf{0.882} & 0.878 & & Attack-8 & \textbf{0.939} & 0.938 & 0.938 & 0.937 \\
& Attack-9 & 0.909 & 0.914 & \textbf{0.915} & 0.914  & & Attack-9 &  \textbf{0.939} & 0.938 & 0.939 & 0.938 \\
\midrule
\multirow{6}{*}{DBLP} & Attack-3 &  0.834 & 0.835 & \textbf{0.839} & 0.837  & \multirow{6}{*}{Photo} & Attack-3 &  \textbf{0.892} & 0.871 & 0.863 & 0.875 \\ 
& Attack-4 &   0.873 & \textbf{0.878} & \textbf{0.878} & 0.873 & & Attack-4 & 0.916 & 0.890 & 0.896 & 0.899 \\
& Attack-5 &  0.872 & 0.872 & \textbf{0.878} & 0.874  & & Attack-5 &  \textbf{0.915} & 0.900 & 0.902 & 0.902 \\ 
& Attack-6 & 0.879 & 0.883 & \textbf{0.894} & 0.879 & & Attack-6 &  0.967 & 0.920 & 0.956 & \textbf{0.968}  \\
& Attack-7 & \textbf{0.903} & 0.879 & 0.876 & 0.882  & & Attack-7 &  \textbf{0.968} & 0.909 & 0.952 & 0.966 \\
& Attack-8 & 0.924 &  \textbf{0.927} &  0.926 &  0.917 & & Attack-8 &  \textbf{0.946} & 0.925 & 0.936 & 0.941 \\
& Attack-9 & \textbf{0.929} &  0.926 &  0.925 &  0.921 & & Attack-9 & \textbf{0.946} & 0.927 & 0.935 & 0.943 \\
\midrule
\multirow{6}{*}{CS} & Attack-3 &   0.869 & \textbf{0.872} & 0.850 & 0.843 & \multirow{6}{*}{LastFM} & Attack-3 & \textbf{0.883} & 0.876 & 0.880 & 0.878\\
& Attack-4 & \textbf{0.890} & 0.881 & 0.884 & 0.874 & & Attack-4 & 0.909 & 0.897 & \textbf{0.910} & 0.890 \\
& Attack-5 &  \textbf{0.893} & 0.875 & 0.885 & 0.879 & & Attack-5 & \textbf{0.911} & 0.910 & 0.909 & 0.907 \\
& Attack-6 &  \textbf{0.955} & 0.913 & 0.917 & 0.935  & & Attack-6 & \textbf{0.919} & 0.891 & 0.911 & 0.886 \\
& Attack-7 & \textbf{0.932} & 0.919 & 0.921 & 0.929  & & Attack-7 &  \textbf{0.921} & 0.899 & 0.913 & 0.874 \\
& Attack-8 & \textbf{0.945} &  0.940 &  0.943 &  0.939 & & Attack-8 & \textbf{0.929} & 0.924 & 0.923 & 0.914 \\
& Attack-9 & \textbf{0.940} & 0.936 & \textbf{0.940} & 0.935 & & Attack-9 & \textbf{0.930} & 0.927 & 0.927 & 0.912 \\
\bottomrule
\end{tabular}
}
\end{table*}

\end{document}